%% file: diag.tex
\documentclass[12pt]{iopart}
\include{defrys}

\include{defsym}
\usepackage{latexsym,epsfig}
\usepackage{savesym}
\usepackage{iopams}
\usepackage{float}
\usepackage{amsthm,empheq}
\usepackage{graphpap}
\usepackage{graphics}

\newcommand{\bF}{\boldsymbol{\mathcal{F}}}
\newcommand{\bT}{\boldsymbol{\mathcal{T}}}
\newcommand{\SC}{\Lambda}
\usepackage{multicol}

\begin{document}

\title{Diagrammatic approach to response problems in composite systems}

\date{\today}

\author{P. Szymczak$^1$ and B. Cichocki}
\ead{$^1$ Piotr.Szymczak@fuw.edu.pl}
\address{Institute of Theoretical Physics, Warsaw University,
 Ho\.za 69, 00-681 Warsaw, Poland}

\begin{abstract}
The bulk macroscopic response of a system of particles or inclusions
with field-induced forces is studied. The susceptibilities and transport
coefficients in such a system are expressed as averages of a multiple
scattering expansion. A special diagrammatic method is developed to analyze the
structure of the expansion. The concept of irreducibility is discussed in detail and shown
to be crucial in obtaining macroscopic equations characterizing the system response with
coefficients depending solely on local properties of the medium. Due to the representation
of particles by lines in diagrams, irreducibility is given a particularly simple
topological interpretation in the diagrammatic language. The method is illustrated by a
discussion of response problems in colloidal suspensions in presence of hydrodynamic
interactions.
 \end{abstract}

\maketitle


\section{Introduction}

\eqnobysec
Calculation of the effective properties of a composite system is an active field of
research (see e.g. \cite{Torquato:2002,Sahimi:2003,Landauer:1978} and references therein)
important not only for the physical insight it provides but also for many potential
practical applications. The composite materials considered here are systems of particles
or
inclusions embedded in a homogeneous medium and subject to an external field.
A classical example of such a system is the Kirkwood-Yvon dielectric \cite{
Kirkwood:1936,Ramshaw:1972} - a set of polarizable, spherical inclusions embedded in a
uniform and isotropic medium. Its relative simplicity makes it a convenient starting point
to illustrate the methods presented here. Next, we focus on a more complicated composite
system - a colloidal suspension, in which the motion of suspended particles in the liquid
is caused either by the gravitational force or by an imposed external flow.

To define the effective macroscopic properties, one must start from local equations that
govern the system response to external disturbances. The construction of such equations is
not trivial if there are long-range interactions present, since they often lead to the
divergent integrals in the expressions for the transport coefficients. Those divergences
are
usually removed with use of rather subtle ``regularization'' techniques (e.g.
\cite{Bedeaux-Mazur:1973,FelderhofFC:1982,Beenaker:1984,Cichocki-Felderhof:1988:1}),
which involve nontrivial manipulation of multiple scattering expansion
with the careful resummation of the various kinds of terms. The calculation may be
facilitated by the development of diagrammatic methods which not only allow the local
response equations to be obtained quickly and reliably, but also provide us with a clear
interpretation of the different steps in the regularization procedure, which are sometimes
obscured in the standard approach.

A key factor for a successful diagrammatic method is the requirement that the structure of
the terms of the scattering expansion should be reflected in topological properties of
respective diagrams. In particular, since a given particle may take part in more than one
scattering event, it is convenient to represent particles  not by points
but by lines in analogy to the diagrammatic techniques developed by the Brussels group
\cite{Balescu:1975} in nonequilibrium statistical physics.
In particular, due to the representation of the particles
by lines in our diagrammatic approach, a natural
ordering of the successive scattering events in the multiple scattering expansion is
reflected in the ordering of the scattering events along the particle line.
Additionally, the notion of irreducibility, central to the
regularization procedure, is now given an elegant interpretation
in terms of the topology of the diagrams.
This constitutes a fundamental difference between our approach and
another diagrammatic technique found in the literature, due to
Barrera~\cite{Barrera:1989,Barrera:1992}. In Barrera approach the particles are
represented
by points, which complicates the analysis, since the diagrams then become multiply-
connected, i.e. there is usually more than one edge linking the nodes. Hence the
edges must be numbered in order to obtain a unique identification for a particular
diagram.
This makes it harder to analyze various types of diagrams and to link the structure of the
multiple scattering expansion to the their topological properties.
Additionally, there is no obvious generalization of that technique to the
time-dependent case, in contrast to the particle line approach.

The regularization procedure with use of the diagrammatic technique allows one to obtain
well-defined theoretical expressions for transport coefficients, free of the integrals
diverging with the size of the system, even in the presence of long-range forces.
In particular, as it will be shown in a subsequent paper, the diagrammatic expansion
allowed us to construct a hierarchy of equations for the correlation functions in
a settling suspension, which in turn allows to solve the long-standing problem
of velocity fluctuations in non-Brownian suspension~\cite{SadlejCichocki}. Namely, it was
argued theoretically more than 20 years ago by Caflisch and Luke \cite{CaflischLuke1985}
that the velocity fluctuations
should diverge linearly with the macroscopic size of the system. However, this prediction
has not been confirmed in the experiments~\cite{Nicolai1995,Nicolai1995b,Segre1997}.
Instead, in most of the experiments, the saturation of the velocity fluctuations was
observed. A careful analysis of the correlation structure of the settling suspension, in
which the diagramatic analysis plays a fundamental role, allowed us to show that the
velocity fluctuations do not diverge with increasing container dimensions.
Another problem of a similar nature is the calculation of the mean velocity of a settling
non-Brownian suspension. Batchelor \cite{Batchelor1982,BatchelorWen1982} calculated this
quantity for the polydisperse suspension. It turns out, however, that his theory gives
ambiguous results for the monodisperse case (the result depends on the way the limit is
taken) \cite{feu1991}. Also in this case, a diagrammatic analysis allows us to derive a
well-defined and unambiguous result for the sedimentation velocity in both polydisperse
and monodisperse case~\cite{SadlejCichocki}.

The diagrammatic expansion constitutes also a good starting point for
the construction of various approximation
methods for calculating the effective properties of the medium.
In general, transport coefficients have different values in
the short-time regime i.e., for times in which
particles have hardly moved and for long times
when the relaxation of the distribution of particle positions
becomes important. This relaxation gives rise to the memory
effects, which can also be incorporated into the presented diagrammatic approach.
Additionally, we discuss the relation of our approach to another method of obtaining the
transport coefficients, based on the Fourier space formulation of response equations and
subsequent calculation of the small wavenumber, ${\bf k}\rightarrow 0$, limit of the
response kernels.

\section{Multiple scattering expansion}

A composite medium is often modeled by a disordered system of particles or inclusions
embedded in a homogeneous matrix.
In many cases, if such a system is inserted into the field $\Psi_0(\rr)$, the particles
themselves become sources of the field (as it is the case for polarizable dipole
systems). The contribution of the induced sources to the total field in the sample,
$\Psi(\rr)$ is then given by
\be
\int \de \rr' \gr(\rr,\rr') s(\rr')
\ee
where the function $s(\rr)$ describes the intensity of the sources and $\gr(\rr,\rr')$ is
the Green's function.
The response of the particle $i$ to the field $\Psi_{ext}$, external to the particle,  is
characterized by the operator $\M$
\be
s_i(\rr)=\int \de \rr' \M(i;\rr,\rr') \Psi_{ext}(\rr'), \ \ \ \ \ i=1,\dots,
N
\label{source}
\ee
with
\be
M(i;\rr,\rr')=\theta_i(\rr) M(i;\rr,\rr') \theta_i(\rr')
\label{tmt}
\ee
where $\theta_i$ is a
characteristic function of $i$th particle.
The above equation reflects the fact that the induced sources
$s_i$ depend only on the values of the field $ \Psi_{ext}(\rr')$ within the particle $i$
and vanish outside the particle.

These ideas may be illustrated with an example of the Kirkwood-Yvon
dielectric \cite{Kirkwood:1936,Ramshaw:1972} - a system
of  $N$ identical polarizable point dipoles. In this case the field $\Psi$ corresponds
to the electric field in the dielectric whereas the sources
${\bf s}_i(\rr)$ are expressed in terms of the dipole moments, ${\bf p}_i$, as
\be
{\bf s}_i(\rr) = {\bf p}_i \delta(\rr-\rr_i)
\ee
The Green's function is then given by dipole-dipole interaction tensor
\begin{equation}
\begin{split}
& \gr(\rr,\rr')= \boldsymbol{\Hat{\cal G}}(\rr-\rr'), \\
& \boldsymbol{\Hat{\cal G}}(\rr) = \nabla \nabla \frac{1}{r} = - \frac{
\boldsymbol{1}}{r^3}
+
\frac{3 \rrh \rrh}{r^3}.
\label{te}
\end{split}
\end{equation}
Finally, the single particle scattering operator is simply
\be
M(i;\rr,\rr')=\delta(\rr-\R_i) \alpha \delta(\rr'-\R_i)
\ee
where $\alpha$ is the molecular polarizability and $\R_i$ - position of $i$
th dipole.

Let us find the response of a composite system to the external field
 $\Psi_0(\rr)$. The total field in the sample is then given by
\be
\Psi=\Psi_0 + \sum_i \gr s_i
\label{field}
\ee
whereas
\be
s_i = \M(i) \left( \Psi_0 + \gr \sum_{j \neq i} s_j \right)
\label{rel}
\ee
In the above, the shorthand notation is used, in which the integrations and the
coordinates ($\rr, \rr'$) are suppressed,
i.e.
\be
(AB) (\rr) \equiv \int A(\rr,\rr') B(\rr') d\rr'
\ee
Additionally, the space arguments of the operators ($\rr, \rr'$ etc.) are dropped.
Note that the term $i=j$ is omitted in the summation \eqref{rel} since the response
relation \eqref{source} relates the sources $s_i$ to the field external with respect to
the
particle $i$.

The relation \eqref{rel} is iterated to obtain successive terms of a
multiple-scattering
expansion
\be
s_i =( \M(i) + \sum_{j \neq i} \M(i) \gr \M(j) + \dots ) \Psi_0.
\label{itt}
\ee
Using the above formalism, one can find the response kernel $T$ defined by
the relation
\be
s = T \Psi_0,
\label{defT}
\ee
where $s$ is the total source intensity
\be
s=\sum_i s_i.
\ee
Using~\eqref{itt} one may represent $T$ in the form of a scattering
expansion
\be
T= \sum_i \M(i) + \sum_i \sum_{j \neq i} \M(i) \gr \M(j) + \dots
\label{tt}
\ee

\section{Averaging the scattering expansion over configurations}

Since we are interested in the average behaviour of the system on a macroscopic level, its
response should be averaged over an ensemble of particle configurations. Averaging of~
\eqref{tt} leads to
\begin{equation}
<s> = <T> \Psi_0
\label{tta}
\end{equation}
where the brackets stand for a configurational average
\begin{equation}
<A>=\int A(\X,\rr,\rr') P(\X) d \X,
\label{calka}
\end{equation}
and $P(\X)$ is the configurational probability distribution function, with
$\X=\{\R_1, \R_2, \dots, \R_N\}$.

In the dielectric example considered above this corresponds to the relation between the
external electric field, ${\bf E}_0$, and the polarization, ${\bf P}$,
\be
{\bf P} = <{\bf s}> =  <T> {\bf E}_0.
\ee
However, the above relation is not local, since polarization in the sample depends not
only
on ${\bf E}_0$ but also on the shape of the sample, boundary conditions etc. Conversely, a
local relation characterizing dielectric response is
\be
{\bf P} = \epsilon_0 \chi <{\bf E}>
\ee
where $<{\bf E}>$ is the macroscopic electric field. The electric susceptibility $\chi$
does not depend on the shape or size of the sample but only on the local properties of the
material. In particular, the dielectric constant of a medium is expressed as
\be
\epsilon = 1 + \chi
\ee
Going back to the general case, we see that the operator $<T>$ may not be a physically
meaningful measure of system's response. Instead, one should study
the response operator $X$ defined by the relation
\begin{equation}
<s> = X <\Psi>
\label{xx}
\end{equation}
linking the sources $<s>$ to the total field inside the sample, $<\Psi>$. The procedure of
obtaining $X$ from $<T>$ (so-called ``reduction'' or ``regularization'' of a response
kernel) is presented below
in a systematic way.

First, we represent the operator $T$ as

\begin{equation}
T=\sum_{i} T(i) + \frac{1}{2!} \sum_{i \neq j} T(i,j) + \frac{1}{3!} \sum_{i
\neq j \neq k}
T
(i,j,k) + ... ,
\label{grup}
\end{equation}

\ii where $T(i_1,...,i_s)$ comprises all these terms in the scattering
sequence in which
all the  particles $\{i_1,i_2...i_s\}$ are included.

Hence we get for $<T>$
\be
 <T> =  \sum_{s=1}^{N} \frac{N!}{(N-s)! s!} \int T(1,2...,s) P(1, \dots, N)
\mbox{d} 1 ... \mbox{d} N,
\ee
where we write $i$ instead of $\R_i$ to simplify notation. The above expression may also
be
written as
\be
 <T> = \sum_{s=1}^{N} \frac{1}{s!} \int T(1,2...,s) n(1,2,\dots,s)
\mbox{d} 1 ... \mbox{d} s,
\label{zen}
\ee
where $n(1,2...,s)$ is the s-particle partial distribution function
\begin{equation}
n(1,2,...,s) = \frac{N!}{(N-s)!} \int P (1,..,N) \mbox{d} (s+1) ... \mbox{d} N.
\end{equation}
Note that the s-particle partial distribution function can be written as
\begin{equation}
 n(\rr_1,\rr_2,...\rr_s)= < \sum_{i_1,i_2,\dots,i_s} \!\!\!\!\! {}^{{}^{\displaystyle
\prime}}
\delta(\rr_1-\R_{i_1}) \delta(\rr_2-\R_{i_2}) ... \delta(\rr_s-\R_{i_s}) >,
\label{mikro}
\end{equation}
which in a shorthand notation will be also denoted as
$ <1  \ 2 \  ... \ s > $. The sum
$\sum'$ in the above expression is supplied with the condition that all
$i_k, \ k=1,\dots,s$
are different each from the other. The above definition \eqref{mikro} of
partial distribution function holds also for a system with a variable number of
particles (if the grand canonical ensemble is used). In this
case the sum in \eqref{zen} should be extended up to infinity:
$\sum_{i=1}^{\infty}$ \cite{Hansen-McDonald:1976}.

Next we assume that the correlations between the two groups of particles
vanish as the
distance between them goes to infinity. This means that the partial
distribution function
should have the group property, i.e.
\begin{equation}
n(1,2...,r,r+1,..,s) \rightarrow n(1,2...,r) n(r+1,...,s),
\end{equation}
as the distance between the particles $\{1,2...r\}$ and $\{r+1,...,s\}$ goes
to infinity.

This property of the partial distribution function allows us to decompose
$n(1,2...s)$ as
\cite{Balescu:1975}

 \begin{align}
 & n(1) = \h(1), \nonumber \\
 & n(1,2) = n(1)n(2) + \h(1,2), \nonumber \\
 & n(1,2,3) = n(1)n(2)n(3) + n(1) \h(2,3) + n(2) \h(1,3) + n(3) \h(1,2) + \h
(1,2,3), \no \\
 & ... ,
\label{clusn}
\end{align}
where the s-particle correlation function $h(1,2...s)$ which vanishes whenever any subset
of particles $\subset$ $\{1,2...s\}$ is dragged away from the rest.

With the above decompositions one can write the average $<A>$ in \eqref{calka} as the sum
of terms of the general form
\be
T_s(\SC,c) = \int \SC(i_1,i_2,\dots,i_s) c(i_1,i_2,\dots,i_s) \mbox{d}i_1 ... \mbox
{d}i_s,
\label{tcs}
\ee
where $c(i_1,i_2,\dots,i_s)$ is a product of a number of correlation functions involving
particles
$\{i_1,\dots,i_s\}$ whereas $\SC(i_1,i_2,\dots,i_s)$ is one of the scattering sequences
making up $T(i_1,i_2,\dots,i_s)$.

For the later use we introduce after Michels \cite{Michels:1989} the ``uncorrelating
operator"
\begin{equation}
P_{unc} = ><,
\label{unc}
\end{equation}
which has the property of statistically uncorrelating the variables at its
left from those
at
its right, i.e.
\begin{equation}
< A P_{unc} B> = <A><B>.
\end{equation}
The orthogonal complement of $P_{unc}$ is
\begin{equation}
Q_{unc} = {\bf 1} - ><.
\end{equation}
So, using the notation of Eq. (\ref{mikro}), we get for example
\begin{equation}
<1 \ Q_{unc} \ 2> = <1 \ 2>-<1><2>=n(1,2)-n(1)n(2) = \h(1,2).
\end{equation}

The decomposition (\ref{clusn}) together with the cluster expansion (\ref{grup}) leads to
the representation of response kernels as
sums of many-body terms from the scattering sequence multiplied by
respective correlation functions. To deal effectively with such a
complicated structure a special diagrammatic technique is employed.

\section{Diagrammatic representation}

We introduce the diagrammatic representation of the scattering (S)
and correlation (C) structure of the kernels. Such SC diagrams consist of
the following
elements

\begin{enumerate}
\item the horizontal line - - - - represents a given particle (also called
particle line)
\item the symbol \  \akd \ \ \ stands for the operator $\M(i)$
\item the vertical line $\mid$ stands for the $\gr$ - bond
\item double vertical line represents
the correlation function h (called h-bond)
\end{enumerate}

The exact interpretation of an h-bond depends on the geometric structure of
a diagram. For
example

\begin{picture}(120,60)

\put(49,48){$\boldsymbol{\circ}$}

\put(49,28){$\boldsymbol{\circ}$}

\put(49,8){$\boldsymbol{\circ}$}

\put(43,48){1}

\put(43,28){2}

\put(43,8){3}

\put(51,12){\line(0,1){37}}

\put(53,12){\line(0,1){37}}

\end{picture}

stands for $h(1,2,3)$, whereas

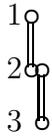
\begin{picture}(120,60)

\put(49,48){$\boldsymbol{\circ}$}

\put(49,28){$\boldsymbol{\circ}$}

\put(53,28){$\boldsymbol{\circ}$}

\put(53,8){$\boldsymbol{\circ}$}

\put(43,48){1}

\put(43,28){2}

\put(43,8){3}

\put(55,12){\line(0,1){17}}

\put(57,12){\line(0,1){17}}

\put(51,32){\line(0,1){17}}

\put(53,32){\line(0,1){17}}

\end{picture}

\ii corresponds to $h(1,2) h(2,3)$.

Moreover, if the first symbol on the particle line (looking form the
left side) is
filled,
then the position of this particle is integrated over. Hence, for example
the diagram
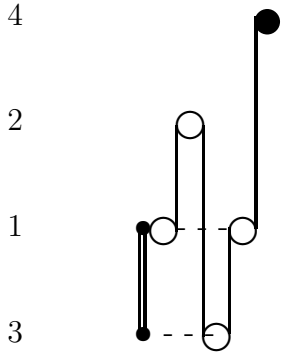
\begin{picture}(120,160)

\put(57,50){\kdp}

\put(67,90){\kdp}

\put(77,10){\kdp}

\put(87,50){\kd}


\put(88,123){\Huge $\bullet$}

\thicklines

\put(60,50){\line(0,1){40}}

\put(70,90){\line(0,-1){80}}

\put(80,11){\line(0,1){40}}

\put(90,51){\line(0,1){80}}

\put(46,12){\line(0,1){37}}

\put(48,12){\line(0,1){37}}

\put(44,48){$\boldsymbol{\bullet}$}

\put(44,8){$\boldsymbol{\bullet}$}


\put(60,48){- - -}

\put(55,8){- - -}


\put(-4,8){3}

\put(-4,48){1}

\put(-4,88){2}

\put(-4,128){4}

\put(420,28){(D 1)}

\end{picture}


\ii represents the expression

 \begin{align}
\int \de 1 \de 3  \de 4 \, & h(13) \M(1) \gr(12) \M(2) \gr(23)
\M(3) \gr(31) \M(1) \gr(14) \M(4).
\label{krotki}
\end{align}

Note that the diagrams should be
read from left to right. The particles lines $2$ and $4$ in the above diagram are left out
since there's only a single operator involving each of these particles.

\subsection{Irreducibility}

A key notion in the analysis of internal structure of scattering sequence terms is
the concept of {\bf irreducibility} of a diagram.
Namely, the $\gr$ bond in the diagram is called a connection line if the removal of this
$\gr$-bond causes the diagram to become disconnected. Diagrams with one or more connection
lines are called {\bf reducible}, whereas diagrams without any connection lines - {\bf
irreducible}.

For example the diagram

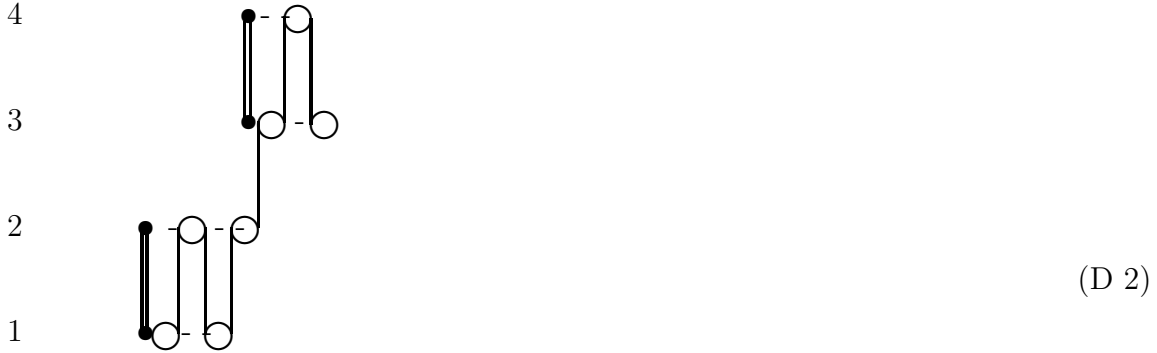
\begin{picture}(120,150)

\put(57,10){\kdp}

\put(67,50){\kdp}

\put(77,10){\kd}

\put(87,50){\kd}

\put(97,90){\kdp}

\put(107,130){\kdp}

\put(117,90){\kdp}

\thicklines

\put(60,11){\line(0,1){40}}

\put(70,51){\line(0,-1){40}}

\put(80,11){\line(0,1){40}}

\put(90,51){\line(0,1){40}}

\put(100,91){\line(0,1){40}}

\put(110,130){\line(0,-1){40}}

\put(46,12){\line(0,1){37}}

\put(48,12){\line(0,1){37}}

\put(85,92){\line(0,1){37}}

\put(87,92){\line(0,1){37}}

\put(56,48){- \, \ - -}

\put(61,8) {- -}

\put(100,88){ - }

\put(87,128){ - - }

\put(-5,48){2}

\put(-5,8) {1}

\put(-5,88){3}

\put(-5,128){4}

\put(44,48){$\boldsymbol{\bullet}$}

\put(44,8){$\boldsymbol{\bullet}$}

\put(83,88){$\boldsymbol{\bullet}$}

\put(83,128){$\boldsymbol{\bullet}$}

\put(400,28){(D 2)}

\end{picture}




\ii is reducible and can be cut into two pieces by breaking the $\gr$
- bond between  particles 2 and 3. The connection line which
is most to the left will be called {\bf articulation line}.
Note that the sub-diagram on the left of the articulation line is
irreducible.

In the analogous way we can define the reducibility for the scattering structure of
the diagrams (S-reducibility). First of all {\bf nodal line} is defined as a $\gr$-bond
which would be a connection line if all the $h-bonds$ in a diagram are removed. Diagrams
with one or more nodal
lines are called S-reducible.

Hence in the following diagram


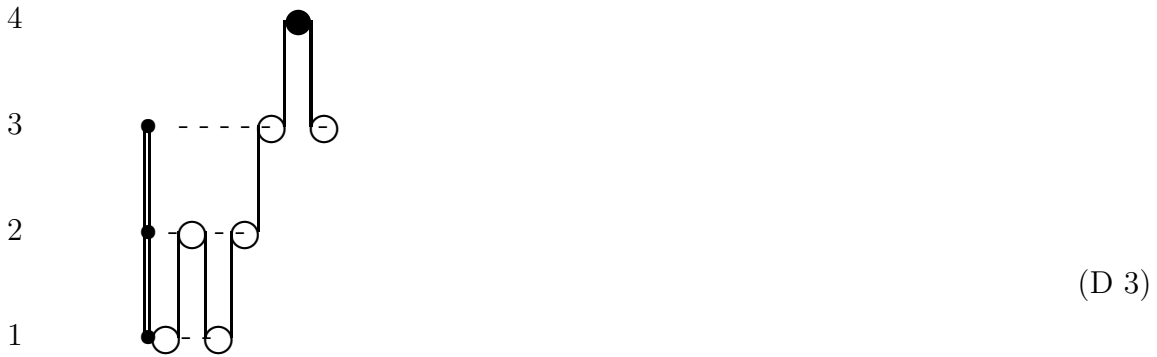
\begin{picture}(120,150)

\put(57,10){\kdp}

\put(67,50){\kdp}

\put(77,10){\kd}

\put(87,50){\kd}

\put(97,90){\kdp}

\put(99,124){\Huge $\bullet$}


\put(117,90){\kdp}

\thicklines

\put(60,11){\line(0,1){40}}

\put(70,51){\line(0,-1){40}}

\put(80,11){\line(0,1){40}}

\put(90,51){\line(0,1){40}}

\put(100,91){\line(0,1){40}}

\put(110,131){\line(0,-1){40}}

\put(47,12){\line(0,1){37}}

\put(49,12){\line(0,1){37}}

\put(47,52){\line(0,1){37}}

\put(49,52){\line(0,1){37}}

\put(56,48){- \, \ - -}

\put(61,8) {- -}

\put(56,88){ - - - - - \ \, \ -}

\put(-5,48){2}

\put(-5,8) {1}

\put(-5,88){3}

\put(-5,128){4}

\put(45,48){$\boldsymbol{\bullet}$}

\put(45,8){$\boldsymbol{\bullet}$}

\put(45,88){$\boldsymbol{\bullet}$}

\put(400,28){(D 3)}

\end{picture}




\ii the $\gr$ bond between particles 2 and 3 is the nodal line but
not the connection line and the diagram is S-reducible, although it is
irreducible with respect to its full SC-structure (which includes both correlation
and scattering part).

\subsection{The nodal structure}

The nodal lines decompose the particles in a given diagram on the set of {
\bf nodal blocks}
$C_i$:
$C_1$ denotes the set of particles on the left of the first nodal line,
$C_2$ - the
particles
between the first and the second nodal line and so on. Note that the
definition of the nodal
line assures that $C_i \cap C_j = \emptyset$ if only $i \neq j$.

For example the diagram

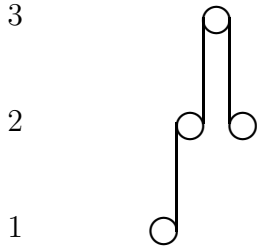
\begin{picture}(120,150)

\put(57,50){\kdp}

\put(67,90){\kd}

\put(77,130){\kd}

\put(87,90){\kdp}

\thicklines

\put(60,50){\line(0,1){41}}

\put(70,91){\line(0,1){40}}

\put(80,131){\line(0,-1){40}}

\put(-4,128){3}

\put(-4,88){2}

\put(-4,48){1}

\put(400,28){(D 4)}

\end{picture}



\ii has the nodal structure of the form


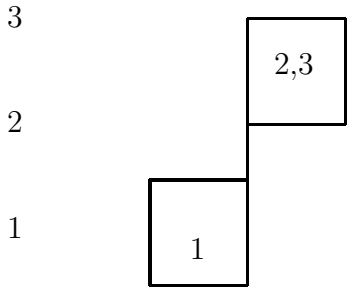
\begin{picture}(120,180)

\thicklines

\put(77,30){\line(0,1){91}}

\put(77,91){\line(0,1){40}}

\put(114,131){\line(0,-1){40}}

\put(77,91){\line(1,0){37}}

\put(114,131){\line(-1,0){37}}

\put(77,30){\line(-1,0){37}}

\put(40,30){\line(0,1){40}}

\put(40,70){\line(1,0){37}}

\put(87,110){2,3}

\put(55,40){1}

\put(-14,128){3}

\put(-14,88){2}

\put(-14,48){1}

\end{picture}



\ii or simply $1|23$.

The structure in the above figure is called the {\bf nodal structure graph}
(NSG).
The vertices of such a graph are nodal blocks, whereas the bonds in this
graph are created
by
nodal lines.

\subsection{The block distribution function}

Consider all the irreducible diagrams which have the same scattering
structure and differ only in correlation structure. The task of summing all
of these diagrams thus boils down to finding the sum of all their correlation functions.

To start with, the condition of irreducibility requires that if there is a nodal line in
the diagram then
particles on the left of it cannot be totally uncorrelated from particles on
its right. This
means that
the correlation function that we are looking for is given by
 \begin{equation}
b(C_1|C_2|...|C_k) = < C_1 (1-P_{unc}) C_2 (1-P_{unc}) ... (1-P_{unc}) C_k
>.
\label{bb}
\end{equation}
\ii Here $C_1|C_2|...|C_k$ describes the nodal structure of the diagram,
whereas
the operator $P_{unc}$ is the "uncorrelating operator" introduced in
(\ref{unc}).
The function $b(C_1|C_2|...|C_k)$ defined in (\ref{bb}) is called the block
distribution
function \cite{FelderhofFC:1982}.
Note that if there are no nodal lines in the scattering structure of a given
s-particle
diagram, than $b$ would be just the full s-particle partial distribution
function $n(1,2,...,s)$.

To get a better grip on $b(C_1|...|C_k)$, let us evaluate it for a few simple
scattering sequences. For the sequence presented in Diagram (D 4)
the block distribution reads:
\begin{equation}
b(1|23) = <1 (1-P_{unc}) 23> = <123> - <1><23> = n(1,2,3) - n(1) n(23).
\end{equation}
We see that $b(1|23)$ goes to zero as the particle $1$ is dragged away
from the particles $2$ and $3$, as in this case
\begin{equation}
n(1,2,3) \rightarrow n(1) n(23).
\end{equation}
Let us consider now the scattering sequence of the form


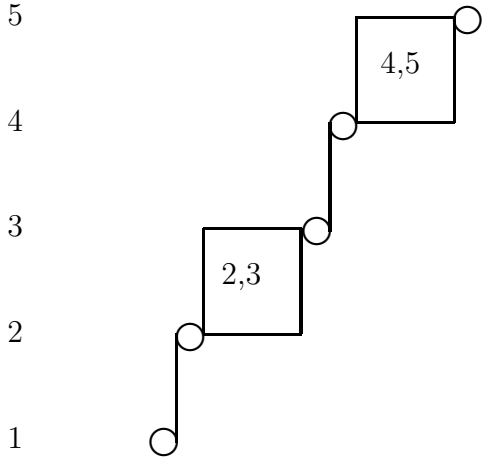
\begin{picture}(120,280)

\put(57,50){\kdp}

\put(67,90){\kd}

\put(115,130){\kd}

\put(125,170){\kd}

\put(172,210){\kdp}

\thicklines

\put(60,50){\line(0,1){41}}

\put(70,91){\line(0,1){40}}

\put(107,131){\line(0,-1){40}}

\put(70,91){\line(1,0){37}}

\put(107,131){\line(-1,0){37}}

\put(118,130){\line(0,1){41}}

\put(128,171){\line(0,1){40}}

\put(165,211){\line(0,-1){40}}

\put(128,171){\line(1,0){37}}

\put(165,211){\line(-1,0){37}}

\put(77,110){2,3}

\put(137,190){4,5}

\put(-4,128){3}

\put(-4,88){2}

\put(-4,48){1}

\put(-4,168){4}

\put(-4,208){5}

\end{picture}


where

\begin{picture}(120,80)

\thicklines

\put(77,11){\line(1,0){137}}

\put(77,51){\line(1,0){137}}

\put(77,11){\line(0,1){40}}

\put(214,51){\line(0,-1){40}}

\put(117,30){$i_1, i_2 \dots i_k$}

\end{picture}




\ii stands for any irreducible scattering sequence that involves the
particles
$i_1,i_2 \dots i_k$.

The above scattering sequence has the nodal structure $(1|23|45)$. Therefore
its block
distribution function reads
 \begin{align}
 & b(1|23|45)=<1 (1-P_{unc}) 23 (1-P_{unc}) 45 > =  \\
 & =  <12345> - <1><2345>
- <123><45> + <1><23><45> = \no \\
 & = n(1,2,3,4,5) - n(1)n(2,3,4,5) - n(1,2,3) n(4,5) + n(1)n(2,3)n(4,5), \no
\end{align}
\ii which, as can be easily proved, vanishes whenever the particle $\{1\}$
is separated from the rest
or the group
$\{1,2,3\}$ is dragged away from $\{4,5\}$.

\section{Reduction of the diagrams}

\label{redi}

In Section II we have obtained the representation of the response kernel
$<T>$  as the sum of terms of the form \eqref{tcs}.
Each such term may be represented as a diagram, according to the
rules formulated above.
Next, those diagrams may be divided into two groups: reducible and
irreducible ones.
Thus
$$
<T>  = <T>^{irr} + <D>
$$
where $<T^{irr}>$ is the sum of all irreducible diagrams of $<T>$ whereas
$<D>$ - is the sum
of the reducible ones.
However, each reducible diagram may be written in form of a product:
\begin{equation}
D  = I \gr R
\label{to}
\end{equation}
\ii where $D$ stands for the diagram under consideration, $I$ is its part to
the left of the
articulation
line and $R$ is the part to the right of the articulation line. As follows from the
definition of irreducibility, the diagram corresponding to $I$ must be
irreducible, since it does not contain an articulation line itself.  For example, the
diagram (D 2) is divided in a following way

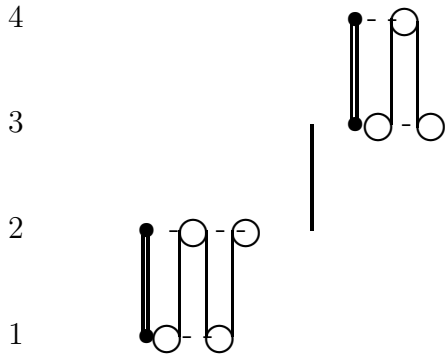
\begin{picture}(120,160)

\put(57,10){\kdp}

\put(67,50){\kdp}

\put(77,10){\kd}

\put(87,50){\kd}

\put(137,90){\kdp}

\put(147,130){\kdp}

\put(157,90){\kdp}

\thicklines

\put(60,11){\line(0,1){40}}

\put(70,51){\line(0,-1){40}}

\put(80,11){\line(0,1){40}}

\put(110,51){\line(0,1){40}}

\put(140,91){\line(0,1){40}}

\put(150,130){\line(0,-1){40}}

\put(46,12){\line(0,1){37}}

\put(48,12){\line(0,1){37}}

\put(125,92){\line(0,1){37}}

\put(127,92){\line(0,1){37}}

\put(56,48){- \, \ - -}

\put(61,8) {- -}

\put(140,88){ - }

\put(127,128){ - - }

\put(-5,48){2}

\put(-5,8) {1}

\put(-5,88){3}

\put(-5,128){4}

\put(44,48){$\boldsymbol{\bullet}$}

\put(44,8){$\boldsymbol{\bullet}$}

\put(123,88){$\boldsymbol{\bullet}$}

\put(123,128){$\boldsymbol{\bullet}$}

\end{picture}




Here $I$ is given by the diagram

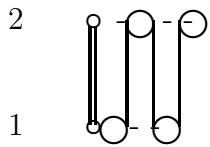
\begin{picture}(120,130)

\put(37,50){\kdp}

\put(47,90){\kdp}

\put(57,50){\kd}

\put(67,90){\kd}

\thicklines

\put(40,51){\line(0,1){40}}

\put(50,91){\line(0,-1){40}}

\put(60,51){\line(0,1){40}}

\put(26,52){\line(0,1){37}}

\put(28,52){\line(0,1){37}}

\put(36,88){- \, \ - -}

\put(41,48) {- -}

\put(-5,88){2}

\put(-5,48) {1}

\put(24,88){$\boldsymbol{\circ}$}

\put(24,48){$\boldsymbol{\circ}$}

\end{picture}




\ii whereas $R$ is given by

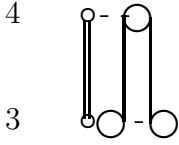
\begin{picture}(20,130)

\put(37,70){\kdp}

\put(47,110){\kdp}

\put(57,70){\kdp}

\thicklines

\put(40,71){\line(0,1){40}}

\put(50,110){\line(0,-1){40}}

\put(25,72){\line(0,1){37}}

\put(27,72){\line(0,1){37}}

\

\put(40,68){ - }

\put(27,108){ - - }

\put(-5,68){3}

\put(-5,108){4}

\put(23,68){$\boldsymbol{\circ}$}

\put(23,108){$\boldsymbol{\circ}$}

\end{picture}


The scattering structure of both $I$ and $R$ diagrams is exactly the same as the
scattering structure of the original $<T>$ diagrams. However,  due to the
irreducibility restriction, the correlation structure of $I$ diagrams is different:
the correlation function which multiplies a sum of
all $I$ diagrams with the given scattering structure is given by the block correlation
function $b(C_1|...|C_k)$ defined in (\ref{bb}). Thus the sum of all $R$
diagrams is just $<T>$, whereas the sum of all $I$ diagrams is $<T>^{irr}$.
These arguments lead to
\be
<T>=<T>^{irr} + <T>^{irr} \gr <T>.
\label{tt2}
\ee
which becomes exact in a thermodynamic limit
\cite{Uhlenbeck-Ford:1962}.
Applying both sides of the above equation to $<\Psi_0>$ and using
\eqref{defT} one gets
\be
<s>=<T> \Psi_0 = <T>^{irr} \Psi_0 + <T>^{irr} \gr <T> \Psi_0
\ee
This equation can be combined with the average of \eqref{field}
\be
<\Psi> = \Psi_0 + \gr <s> = \Psi_0 + \gr <T> \Psi_0
\ee
leading to
\be
<s> = <T>^{irr} \Psi_0 + <T>^{irr} (<\Psi> - \Psi_0) = <T>^{irr} <\Psi>
\ee
which links the sources $<s>$ to the local field inside the sample,
$<\Psi>$. Thus the $\X$
operator in Eq.
\eqref{xx}
may be identified with $<T>^{irr}$.

In the following, we consider a more general form of a response kernel,
namely

\be
A=\sum_i \M_o(i) + \sum_i \sum_{j \neq i} \M_<(i) \gr \M_>(j) +
\sum_i \sum_{j \neq i} \sum_{k \neq j} \M_<(i) \gr \M(j) \gr \M_>(k) + \dots
\label{tp}
\ee
which differs from~\eqref{tt} in that it contains the opening operator
$\M_<(i)$, the
closing operator
$\M_>(i)$ and the single-particle operator $\M_o$, which in general are
different from
$\M(i)$.

The reduction procedure for $<A>$ is similar to the one presented above.
However, due to the presence of  $\M_<(i)$ and $\M_>(i)$ in the scattering sequence of A,
the reduction formula is slightly more complex than \eqref{tt2}:
\be
<A> = <A>^{irr} + <A^<>^{irr} \gr <A^>>
\label{ared}
\ee
where the operators $A^<$ and $A^>$ have scattering sequences
\be
A^<=\sum_i \M(i) + \sum_i \sum_{j \neq i} \M_<(i) \gr \M(j) +
\sum_i \sum_{j \neq i} \sum_{k \neq j} \M_<(i) \gr \M(j) \gr \M(k) + \dots
\label{tp1}
\ee
and
\be
A^>=\sum_i \M(i) + \sum_i \sum_{j \neq i} \M(i) \gr \M_>(j) +
\sum_i \sum_{j \neq i} \sum_{k \neq j} \M(i) \gr \M(j) \gr \M_>(k) + \dots
\label{tp1a}
\ee
respectively.

As an example of a response problem described by a general structure~\eqref{tp} we
consider a colloidal suspension - a system of solid particles immersed in a fluid.

\section{Transport phenomena in colloidal suspensions}

The system under consideration consists of N identical
spherical particles of radius $a$ immersed in an incompressible
fluid of shear viscosity $\eta$. The particle Reynolds number is
assumed to be small so that the inertial effects are negligible
and the fluid can be described by Stokes equations. The sources ${\bf s}_i$ are then the
force
density exerted on the fluid by the particles whereas the role of the field $\Psi$ is
played by the fluid velocity field, $\vv(\rr)$.

As it was shown by Mazur and Bedeaux \cite{Mazur-Bedeaux:1974} if the
particles are impenetrable to the flow and the stick boundary conditions
at their surfaces are assumed, then validity of Stokes equations may be
formally extended inside the particles:
\begin{align}
& \eta \nabla^2 \vv - \nabla p+\f_0(\rr)+\f(\rr)=0,  &  &  \label{st1} \\ &  \nabla
\cdot \vv = 0,   & & \label{st2} \\ & \vv(\rr) = \ub_i(\rr) = \U_i + \w_i \times (
\rr-\R_i)  &
\mbox{for} \ \  |\rr-\R_i| \leq a, & \label{vvs} \\ & p(\rr) = 0   & \mbox{for} \ \
|\rr-\R_i| \leq a.  &
\label{Stokes}
\end{align}
Here $\f_0(\rr)$ is an external force density applied to the fluid, such as
gravity. Next,
$\f(\rr)$ is an induced force density localized on the particle surfaces
\cite{Mazur-Bedeaux:1974,Felderhof:1988:2} and $\U_i$ and $\w_i$ are
translational
and rotational velocities of the particles.

The solution of hydrodynamic equations \eqref{st1},\eqref{st2} can be written as
\begin{equation}
\vv(\rr)=\vv_0(\rr)+\int \gr(\rr,\rr') \cdot \f(\rr') d\rr',
\label{flow}
\end{equation}
where $\vv_0(\rr)$ is the flow in absence of the particles and
$\gr(\rr,\rr')$ is the Green
tensor. For an unbounded fluid $\gr(\rr,\rr')$ is given by the Oseen tensor
$\boldsymbol{\gr_0}$
 \begin{align}
& \gr(\rr,\rr')=\gr_0(\rr-\rr'), \no \\
& \gr_0(\rr) \equiv \frac{1}{8 \pi \eta} \frac{\boldsymbol{1} + \rrh
\rrh}{r}, \ \ \ \ \ \ \
\rrh=
\frac{\rr}{r},
\label{ge}
\end{align}

The response of a single particle to the fluid field is described by the
one-particle
friction kernel
$\Z_o(i)$
\be
\f_i(\rr) = \int \Z_o(i;\rr,\rr') (\ub_i(\rr') - \vv_a(\rr')) \de \rr'
\ee
where $\vv_a(\rr)$ is the flow field external to particle $i$. The above equation is
a counterpart of the relation \eqref{source}, with the operator $\M$ corresponding to
$-\Z_o$.
The explicit form of $\Z_o(i)$ for variety of boundary conditions
may be found e.g. in \cite{Cichocki-Felderhof-Schmitz:1988}. Next, we may
proceed in several ways.

In a {\bf friction problem}, one looks for the forces induced on the
particles for the given
flow field. This leads to the relation
\begin{equation}
\f(\rr)=\sum_i \f_i(\rr) = \int \ZZ(\rr,\rr') \cdot (\vv(\rr')-\vv_0(\rr')) \de \rr',
\label{friction}
\end{equation}
where the friction kernel $\ZZ(\rr,\rr')$ can be represented in form of the
scattering
expansion \eqref{tt}
\be
\ZZ= \sum_i \Z_o(i) - \sum_i \sum_{j \neq i} \Z_o(i) \gr \Z_o(j) + \dots
\label{sf0}
\ee
The above relations are analogous to \eqref{defT} and \eqref{tt} respectively. When
deriving Eq.~\eqref{friction}, we used the fact that the operators $\Z_o(i;\rr,\rr')$
are localized inside the corresponding particles, together with the condition \eqref{vvs}.
Additionally, the notation may be simplified further by introducing the operators $\ZZ_o$
and $\G$:
\begin{equation}
{\ZZ_o}_{ij}=\Z_o(i) \delta_{ij} \ \ \ \ \ \ \ \ \ \ \ \G_{ij}=\gr(ij)
(1-\delta_{ij})
\label{sf}
\end{equation}
which are the NxN operator matrices in the particle indices. In the above, $\gr(ij)$
denotes the operator $\gr$ placed between $\Z_o(i)$ and $\Z_o(j)$ in the scattering
expansion \eqref{sf0}. Here and below we use the
script letters ($\ZZ_o$, $\G$, ${\bF}$ \dots) for objects acting in the particle index
space. With the above notation \eqref{sf0} takes form
\be
\ZZ=   \ZZ_o (1+\G \ZZ_o)^{-1}
\ee
The above allows us to find the friction matrix $\zet$ which is defined by the
relation between the
forces and torques acting
on the particles and their velocities (in the absence of external flow)
\be
\boldsymbol{\tilde{\cal F}} =\zet \boldsymbol{\tilde{\cal U}},
\label{tarcie}
\ee
Here
$\boldsymbol{\tilde{\cal F}} = (\boldsymbol{\cal F},\boldsymbol{\cal T})$ is
the
6N-dimensional vector of forces and torques acting on
each of $N$ particles: $(\boldsymbol{\cal F},\boldsymbol{\cal T}) =
(\F_1,\F_2,...,\F_N,\T_1,...,\T_N)$ whereas
$\boldsymbol{\tilde{\cal U}} = ( \boldsymbol{\cal U},\boldsymbol{\it
\Omega})$ is the vector
of translational
and rotational velocities of the particles $\boldsymbol{\tilde{\cal U}} =
(\U_1,...,\U_N,\w_1,...,\w_N)$. The friction matrix, $\zet$, may be similarly decomposed
as
\be
 \zet
 =\left(
\begin{array}{cc}
{\zet}^{tt}&{\zet}^{tr}\\
{\zet}^{rt}&{\zet}^{rr}
\end{array} \right). \no
\ee
The matrices $\zet^{pq}$ ($p,q=t$ or $r$) are the 3Nx3N Cartesian tensors,
and
the superscripts $^t$ and $^r$ correspond to the translational and
the rotational components, respectively.

Subsequent analysis is facilitated by introduction of multipole expansion.
Namely, one represents the force densities and the velocity field around $i$th particle as
the (infinite dimensional) vectors of successive multipoles:
\be
\f_i(\rr) \rightarrow \left( \begin{array}{cc}
{\bF}_i\\
{\bT}_i \\
{\bf S}_i \\
\dots
\end{array}
\right)
\label{force}
\ee
and
\be
\ub_i(\rr)-\vv_0(\rr)  \rightarrow \left( \begin{array}{cc}
\U_i-\vv_0(\R_i) \\
{\bf \Omega}_i-{\boldsymbol{\omega}}(\R_i) \\
{\bf g}_i \\
\dots
\end{array}
\right)
\label{velo}
\ee

In the above, force multipoles are obtained by the following integrations of $\f(\rr)$
\begin{align}
& \F_i = \int \f(\rr) \theta_i(\rr) \de \rr  \label{sily} \\ & \T_i =
\int (\rr-\R_i) \times \f(r) \theta_i(\rr) \de \rr ,  \no \\ & {\bf S}_i =
\int \overbracket{(\rr-\R_i) \f(\rr)} \theta_i(\rr) ,  \no
\end{align}
where
\begin{equation}
\theta_i(\rr)= \theta(a - |\rr-\R_i|)
\end{equation}
is the characteristic function of the particle $i$ and the overbar stands for the
symmetric
and traceless part of the tensor.

On the other hand, velocity multipoles are obtained by the following differentiations:
\be
\boldsymbol{\omega}(\R_i) = \frac{1}{2} (\nabla \times \vv_0)_{\rr=\R_i}
\ee
$$
{\bf g}_i= \frac{1}{2}[\nabla_{\alpha} \vv_{0,\beta}(\rr) + \nabla_{\beta} \vv_{0,
\alpha}(
\rr)]_{\rr=\R_i}
$$

In the multipole notation, the operators $\gr_0$ and $\Z_o$
become matrices. The friction matrix, defined in \eqref{tarcie}, relates the two
lowest velocity multipoles to the two lowest force multipoles. Therefore it can be
obtained
from the multipole matrix $\ZZ$ by the following projection
 \begin{equation}
\zet = \pe \ZZ \pe.
\label{nn1}
\end{equation}
where $\pe=(\pe^t,\pe^r)$ are the projection operators extracting the two lowest moments
from
the velocity (or force) distribution, i.e.
 \begin{equation}
\boldsymbol{\tilde{\cal F}} = \pe \f,
\end{equation}
and
\be
\boldsymbol{\tilde{\cal U}} = \pe \vv
\ee
Subsequently, we will also use the operator $\pe^d$ which gives the third multipole
of the force field, i.e.
\be
{\bf S}_i = \pe_i^d \f_i,
\label{peid}
\ee
and similarly for the velocity field
\be
{\bf g}_i = \pe_i^d \vv_0.
\ee

Let us now find forces acting on particles in
the presence of the ambient flow $\vv_0$. From Eq.~(\ref{friction}) one gets in this
case
  \begin{equation}
 \boldsymbol{\tilde{\cal F}} = \zet \cdot \boldsymbol{\tilde{\cal U}} - \pe
\ZZ \vv_0.
 \label{dlugie}
 \end{equation}
 The above formalism can also be used to solve the mobility problem: finding
velocities of
the
 particles $\boldsymbol{\tilde{\cal
 U}}$ for given forces $\boldsymbol{\tilde{\cal F}}$ and flow $\vv_0$. In
this case, the
relation
 (\ref{dlugie}) gives
 \begin{equation}
 \boldsymbol{\tilde{\cal U}} = \zet^{-1} \boldsymbol{\tilde{\cal F}} + \zet^
{-1} \pe \ZZ
\vv_0
 \equiv \mur \boldsymbol{\tilde{\cal F}} + \Ce \vv_0,
 \label{uu}
 \end{equation}
 which defines the {\bf mobility matrix} $\mur$
  \begin{equation}
 \mur = \zet^{-1}
 \label{mobi}
 \end{equation}
together with the {\bf convection kernel} $\Ce$
  \begin{equation}
 \Ce = \mur \pe \ZZ.
 \end{equation}
The mobility matrix, $\mur$, allows us
to find translational and rotational velocities of particles in terms of
 forces and torques acting on them in the absence of an external flow
 \be
  \left( \begin{array}{cc}
 {\bf U}\\
 {\bf \Omega}
 \end{array}
 \right)
  =\mur
 \left( \begin{array}{cc}
 {\bF}\\
 {\bT}
  \end{array} \right),
 \label{mobility}
 \ee
 \be
  \mur
  =\left(
 \begin{array}{cc}
 {\mur}^{tt}&{\mur}^{tr}\\
 {\mur}^{rt}&{\mur}^{rr}
 \end{array} \right). \no
\ee

Finally, let us consider a problem of finding
the force density $\f$ for given forces
 $\boldsymbol{\tilde{\cal F}} \neq 0$ and ambient flow $\vv_0$. In this
case, from
(\ref{dlugie})
 and (\ref{friction}) we obtain
 \begin{equation}
 \f =  \C \boldsymbol{\tilde{\cal F}} - \zzr \vv_0.
 \label{ff}
 \end{equation}
 where $\C$ is the transpose of $\Ce$ operator
 \begin{equation}
 \C = \ZZ \pe \mur,
 \label{abel}
 \end{equation}
 while the convective friction kernel $\zzr$ \cite{Schmitz-Felderhof:1982}
is given by
  \be
  \zzr = \ZZ - \ZZ \pe \mur \pe \ZZ.
  \label{zzr}
  \ee
   The operator $\gr \zzr$ produces the velocity fields, which are force-free and torque-
free.

 The scattering expansion for the convective friction kernel $\zzr$ is found
to be
 \begin{equation}
 \zzr = \z (1+\G \z)^{-1}=\sum_{k=0}^{\infty} \z (- \G \z)^k,
 \label{zzd}
 \end{equation}
whereas the mobility operator can be written as
\begin{equation}
 \mur =   \mur_o + \mur_o \pe \ZZ_o \frac{1}{1+\G \z} \G \ZZ_o \pe \mur_o=
\mur_o +
\sum_{k=0}^{\infty}  \mur_o \pe
 \ZZ_o (- \G \z)^k \G \ZZ_o \pe \mur_o,
 \label{smu}
 \end{equation}
\ii where
\begin{equation}
 \mur_o = \frac{1}{\pe \ZZ_o \pe}
 \label{jedno}
 \end{equation}
is the one particle mobility matrix whereas $\z$ is one-particle convective
friction matrix, given by the relation analogous to~\eqref{zzr}
\begin{equation}
 \z = \ZZ_o - \ZZ_o \pe \mur_o \pe \ZZ_o.
 \label{daszek}
 \end{equation}
  Since, similarly to the case of the $\zzr$ operator, the velocity fields produced by
$\gr \zr$ are force-free and torque-free, we obtain the relation
\be
\z \pe = \pe \z = 0
\label{touse}
\ee
which will be used in the following.

Note that the scattering expansion \eqref{smu} is of the form \eqref{tp}
with
$\M_<=\mur_o  \ZZ_o$, $\M_>=\ZZ_o
 \mur_o$,
 $\M_o=\mur_o$, and $\M=-\z$. Analogous scattering expansions for the kernels $\Ce$ and
$\C$ introduced above read~
\cite{Felderhof:1988:2}
  \begin{equation}
\C = \ZZ_o \pe \mur_o -  \zzr \G \ZZ_o \pe \mur_o=\sum_{k=0}^{\infty} (- \z
\G)^k \ZZ_o \pe
\mur_o,
 \label{sc}
 \end{equation}
 \begin{equation}
 \Ce = \mur_o \pe \ZZ_o - \mur_o \pe \ZZ_o \G \zzr=\sum_{k=0}^{\infty}
\mur_o \pe \ZZ_o (-
\G \z)^k.
 \label{sce}
 \end{equation}
To obtain the response of the system  on a macroscopic level, we need to average the
above-defined hydrodynamic kernels over an ensemble of particle configurations. Next, the
reduction procedure is carried out, according to the method outlined in
Section~\ref{redi}.
The kernels are reduced analogously to $A$ in Eqs. (\ref{ared}-\ref{tp1a}).
Using the scattering expansions \eqref{sf},\eqref{smu},\eqref{sc},\eqref{sce}  one
obtains
\be
<\mur^{tt}> = <\mur^{tt}>^{irr} + <\Ce^{\,t}>^{irr} \gr <\C^{\,t}>
\label{bou}
\ee
\be
<\Ce> = <\Ce>^{irr} - <\Ce>^{irr} \gr <\zzr>
\label{bou2}
\ee
\be
<\C> = <\C>^{irr} - <\zzr>^{irr} \gr <\C>
\ee
and
\be
<\zzr> = <\zzr>^{irr} - <\zzr>^{irr} \gr <\zzr>
\label{fv2}
\ee
These relations may be used to transform the response equations introduced
in the previous section. For example, if the constant force $\eg$ is
applied to the particles,  by averaging Eq. \eqref{ff} one gets
\be
\begin{split}
& <\f> =  <\C^{\,t}> \eg - <\zzr> \vv_0 = \\
& (<\C^{\,t}>^{irr} - <\zzr>^{irr} \gr <\C^{\,t}>) \eg -
(<\zzr>^{irr} - <\zzr>^{irr} \gr <\zzr>) \vv_0
\end{split}
\ee
The above may be written in the form
\be
 <\f> =  <\C^{\,t}>^{irr} \eg - <\zzr>^{irr} <\vv>
\label{ff2}
\ee
where $<\vv(\rr)>$ is the average velocity of the suspension as a whole
\be
<\vv> =  \vv_0 + \gr <\f>.
\label{average}
\ee
As it is seen from (\ref{st1}-\ref{Stokes}), the suspension velocity field $\vv(\rr)$ has
a
simple interpretation: it is equal to the
fluid velocity if $\rr$ is inside the fluid
and coincides with the rigid body motion wherever $\rr$ lies
inside the particle.

Another quantity of interest is the average velocity of suspended particles
\be
{\bf U}=\frac{1}{N} <\sum_i \U_i >
\ee
which can be obtained by averaging Eq. \eqref{uu}, using the reduction formulae
\eqref{bou}
and \eqref{bou2} and introducing the average suspension velocity according to
\eqref{average}. Such a procedure leads to:
\begin{equation}
<{\bf U}> = \frac{1}{N} (<\sum_{i,j} \mur_{ij}^{tt}>^{irr} \eg + <\sum_i
\Ce^{\,t}_i>^{irr}
<\vv>),
 \label{uu2}
 \end{equation}

\section{Transport coefficients}

\subsection{Sedimentation and diffusion}\label{sedim}

One of the fundamental  problems in the physics of suspensions is
the sedimentation phenomena - i.e. response of a suspension  to a force field,
e.g., gravity.
 The basic quantity here
is the sedimentation
velocity coefficient $K$, the ratio of the average particle velocity $U$ to
the acceleration of the external force field, $E$
\be
K=\frac{U}{E}
\ee
It is important to note that the sedimentation velocity is measured
in the reference frame in which the fluid as a whole is resting, i.e.
$<\vv>=0$.
In this case Eq.~\eqref{uu2} gives
$$
<{\bf U}> =  \frac{1}{N} <\sum_{i,j} \mur_{ij}^{tt}>^{irr}  \eg
$$
For the isotropic system, $<\sum_{ij} \mur_{ij}^{tt}>^{irr}$ is proportional to the unit
tensor and  the sedimentation coefficient may be then expressed  as
\be
K=\frac{1}{3N} \text{Tr} <\sum_{i,j} \mur_{ij}^{tt}>^{irr}
\label{hab}
\ee
Moreover, this allows one also to find the collective diffusion coefficient, which is
connected to $K$ by the relation \cite{Pusey:1991}
\begin{equation}
 D_c =   \frac{k_B T}{\seso}K
 \label{dcs}
\end{equation}

\subsection{Viscosity}

The effective viscosity of a suspension, $\eta_{eff}$ is
obtained from the relation between the average stress of the
system and the average rate of strain
\be
\sigma = 2  \eta \boldsymbol{\cal I} {\bf g}_{eff}
\label{ii2}
\ee
with the effective value of the strain, ${\bf g}_{eff}$, given by
\be
g_{eff}= \frac{1}{2}[\nabla_{\alpha} <\vv>_{\beta} + \nabla_{\beta} <\vv>_{\alpha}]
\ee
In Eq.~\eqref{ii2}, the tensor $\boldsymbol{\cal I}$ is the fourth rank isotropic tensor,
traceless
and symmetric
in its first and last index pairs:
\begin{equation}
\boldsymbol{\cal I}=\frac{1}{2}(\delta_{\alpha \mu}\delta_{\beta \nu}+\delta_{\alpha
\mu}\delta_{\beta
\nu}-\frac{2}{3}
\delta_{\alpha \beta}\delta_{\mu \nu})
\ee

The stress in the suspension has two
components - from the fluid itself and from the force densities on particle surfaces
\cite{Kim-Karilla:1991}, i.e.
\be
{\boldsymbol \sigma} =  {\boldsymbol \sigma}^{fluid} + {\boldsymbol \sigma}^{part}
\label{part}
\ee
with the particle contribution given by the ensemble average of the stresslet
\be
{\boldsymbol \sigma}^{part} = <\sum_i {\bf S}_i \delta(\rr-\R_i)>
\ee
The partition \eqref{part} allows one to write the effective viscosity in the form
$$
\eta_{eff} = \eta + \Delta \eta
$$

To calculate the effective viscosity, let us consider a problem of finding the force
density $\f$ for the given flow
$\vv_0$ in the absence of
 forces, {\it} $\boldsymbol{\tilde{\cal F}} = 0$. This is a special case of
(\ref{ff})
leading to
 \begin{equation}
 \f =  - \zzr \vv_0.
 \label{eff}
 \end{equation}
In particular, in the viscosity problem, one considers a linear  velocity
field of the form
 \begin{equation}
\vv_0 = {\bf g} \cdot \rr
\label{er}
 \end{equation}
(with a symmetric and traceless matrix ${\bf g}$) and looks for the
stresslet, ${\bf S}_i$ of the induced force
The response equation linking the local values of ${\bf g}_i$ with the induced
stresslet
\be
{\bf S}_i=\sum_{j} \mur_{ij}^{dd} {\bf g}_{j}
\label{vis1}
\ee
defines the operator
\begin{equation}
\mur_{ij}^{dd} = \pe_i^d \zzr \pe_j^d
\label{mudd}
\end{equation}
where the projection operator $\pe_i^d$ defined in \eqref{peid} has been used.

The next step is to take the average over the particle configurations.
Eq. \eqref{ff2} gives then
\be
<f> = -<\zzr> \vv_0 = - <\zzr>^{irr} <\vv>
\label{o2a}
\ee
The stresslet may be obtained by acting on the above with the projection operator
$\pe_i^d$. Expanding the flow field in gradients and taking the lowest term
leads to the following relation between stress and strain
as
\be
{\bf \sigma}^{part} = \frac{1}{N} <\sum_{i,j} \mur_{ij}^{dd}>^{irr} {\bf g}_{eff}
\label{visco2}
\ee
where the relation~\eqref{mudd} has been used.

For the isotropic system the average
tensor $<\mu^{dd}>^{irr}$ must be proportional to ${\cal I}$, thus
\begin{equation}
\Delta \eta= \frac{1}{10N}  <\sum_{i,j} \mur_{ij}^{dd}>^{irr}_{\alpha \beta \beta \alpha}
\label{viss}
\end{equation}

\section{Fourier space formulation}
\label{skins}

\subsection{Sedimentation coefficient}

The transport coefficients defined above are often calculated using
Fourier transform.
In the case of the sedimentation coefficient, one starts with the Fourier
transform of Eq.~\eqref{uu},
which in the absence of an external flow reads
\begin{equation}
<{\bf U}(\kb)>=  K(\kb) \eg(\kb),
\label{row2}
\end{equation}
where
\begin{equation}
{\bf U}(\kb) = \frac{1}{N} \sum_i {\bf U}_i e^{i {\kb} \cdot \R_{i}}
\end{equation}
and
\begin{equation}
  K(\kb) =  \hat{\kb} \cdot  <\mur^{tt}(\kb)> \cdot \hat{\kb}
\label{usual}
\end{equation}
is the wavevector-dependent sedimentation coefficient. In the above,
\be
\mur^{tt}(\kb) =
\frac{1}{N} \sum_{ij} \mur_{ij}^{tt}e^{i {\kb} \cdot \R_{ij}}
\ee
The usual sedimentation
coefficient is then obtained as $\kb \rightarrow 0$ limit of \eqref{usual}
\begin{equation}
 K =   \frac{1}{3} \text{Tr} \lim_{\kb \rightarrow 0}  <\mur^{tt}(\kb)>
 \label{dcs2}
\end{equation}
It is important to realize that the limit $\lim_{\kb \rightarrow 0} <\mur^{tt}(\kb)>$
in the above relation cannot be replaced by the $\kb=0$ value of the kernel,
$<\mur^{tt}(\kb=0)>$. This is caused by the presence of long-range hydrodynamic
interactions in the system. Namely, the propagator $\gr(\rr-\rr')$ contains terms which
decay
asymptotically as $|\rr-\rr'|^\gamma$ with $\gamma \leq 3$. While trying to calculate
$\kb=0$
value of the kernels, those long-range terms give rise to diverging integrals.

An alternative way of calculating the sedimentation coefficient would be to start with the
Fourier transform of Eq.~\eqref{uu2}
\begin{equation}
<{\bf U}(\kb)> = <\mur^{tt}(\kb)>^{irr} \eg + <\Ce^{\,t}(\kb)>^{irr} <\vv(\kb)>,
 \label{uu3}
 \end{equation}
with
\be
\Ce(\kb) =
\frac{1}{N} \sum_{i} \int \Ce^{\,t}_{i}(\rr) e^{i {\kb} \cdot (\R_i-\rr)} \de \rr
\ee
and then use the zero net flux condition \cite{Szymczak-Cichocki:2004}
\be
\vv(\kb=0) = 0
\ee
which holds for incompressible fluid placed in an immobile container. This gives
\be
K = \frac{1}{3} \text{Tr} \lim_{\kb \rightarrow 0}  <\mur^{tt}(\kb)>^{irr} =   \frac{1}{3}
\text{Tr}  <\mur^{tt}(\kb=0)>^{irr}
\ee
which is equivalent to \eqref{hab} and does not involve small wavenumber limits, which
makes it much more convenient in calculations. This time the value at $\kb=0$ is well-
defined
since the long-range terms are absent in irreducible kernels \cite{Szymczak-Cichocki:2004}
and thus those kernels are continuous at $\kb=0$.

\subsection{Viscosity}

The Fourier space formalism may be also used to define the viscosity
coefficient.
First, using the Fourier transform of the Oseen tensor
\begin{equation}
\gr(\kb) = \frac{1}{\eta k^2} (\boldsymbol{1} - \hkb \hkb),
\end{equation}
one writes the velocity field in the absence of the particles as
\begin{equation}
 \eta k^2 \vv_0(\kb) = (\boldsymbol{1} - \hkb \hkb) \f_0(\kb).
\end{equation}
The analogous relation between the average flow field in the presence of the
particles, $<\vv>$, and
the external force density, $\f_0$ will then define the wavevector dependent
effective viscosity
function $\eta_{eff}(k)$
\begin{equation}
 \eta_{eff}(k) k^2 <\vv(\kb)> =  (\boldsymbol{1} - \hkb \hkb) \f_0(\kb).
\label{my1}
\end{equation}
Again, the hydrodynamic viscosity coefficient is defined as the long
wavelength limit of $\eta_{eff}(k)$
\be
\eta_{eff} = \lim_{\kb \rightarrow 0} \eta_{eff}(k)
\ee
The function $\eta_{eff}(k)$ may be expressed in terms of the hydrodynamic
kernels defined above.
To this end we note that the flow field in the presence of particles may
equally well be expressed as
\begin{equation}
\eta k^2  <\vv(\kb)> =(\boldsymbol{1} - \hkb \hkb)(<\f(\kb)>+\f_0(\kb)) .
\end{equation}
Inserting  the Fourier transform of Eq.~(\ref{ff}) yields (for the homogeneous system in
the absence of external forces)
\begin{equation}
  \eta k^2 <\vv(\kb)> =
(\boldsymbol{1} - \hkb \hkb) (\f_0(\kb) -
<\zzr(\kb)> \vv_0(\kb)).
\end{equation}
In the above, the Fourier transform of the kernel $\zzr$ is defined as
\begin{equation}
\boldsymbol{\zzr(\kb)} = \int e^{- i \kb \cdot \rr} \  \zzr(\rr-\rr') \ e^{i \kb
\cdot \rr'} \de \rr.
\label{w3}
\end{equation}
where we used the fact that for a homogeneous system $\zzr(\rr,\rr') \equiv
\zzr(\rr-\rr')$.

Next we eliminate $\vv_0(\kb)$, using the identity
\be
\vv_0 = \frac{1}{1- \gr <\zzr>} <\vv>
\ee
obtained by combining Eq. \eqref{average} with Eq. \eqref{eff}.

Finally
\begin{equation}
  \left(\eta k^2 + (\boldsymbol{1} - \hkb \hkb) \frac{1}{1- \gr <\zzr(\kb)>}<\zzr(\kb)>
\right)  <\vv(\kb)> =
  (\boldsymbol{1} - \hkb \hkb) \f_0(\kb).
  \label{my2}
\end{equation}

Comparing Eq. \eqref{my1} with Eq. \eqref{my2} we obtain
\begin{equation}
\Delta \eta = \eta_{eff}-\eta = \lim_{k \rightarrow 0} \frac{1}{2k^2} \left(
(\boldsymbol{1} - \hkb \hkb) \frac{1}{1- \gr <\zzr(\kb)>}<\zzr(\kb)> \right) :  (
\boldsymbol
{1} - \hkb \hkb)
\label{pp}
\end{equation}
The above relation again involves a cumbersome $\kb \rightarrow 0$
limit which cannot be replaced by the corresponding value at $k=0$, not only because of
the
$1/k^2$ term in \eqref{pp} but also since $<\zzr>$ is a
long-range kernel, ill-defined at $k=0$.
However, Eq. \eqref{fv2} gives
\be
<\zzr>^{irr} = \frac{1}{1- \gr <\zzr>}<\zzr>
\ee
thus the relation \eqref{pp} may be rewritten in terms of the irreducible kernel
$<\zzr(\kb)>^{irr}$
\begin{equation}
\Delta \eta = \lim_{\kb \rightarrow 0} \frac{1}{2k^2} \left(
(\boldsymbol{1} - \hkb \hkb) <\zzr(\kb)>^{irr} \right) :  (\boldsymbol{1} - \hkb
\hkb)
\end{equation}
An explicit expression for the above limit may be obtained  by expanding
Eq. \eqref{w3} in $\kb$ and using the fact
that the fields produced by the operator $\zzr$ are force
free and torque-free. Thus the lowest order term in this expansion is $O(\kb^2)$ and
corresponds to
the third multipole (stress-strain) of force and velocity fields as defined in
\eqref{force} and \eqref{velo}.
The coefficient in this term is thus proportional to the right hand side of
Eq.~\eqref{mudd}, and
the proportionality constant may be obtained by isotropy considerations
(a detailed derivation may be found in Ref. \cite{Cichocki-Felderhof-Schmitz:1989}).
Finally:
\begin{equation}
\Delta \eta  = \frac{1}{10N} \lim_{\kb \rightarrow
0}  <\sum_{ij} \mur_{ij}^{dd}
e^{i {\kb} \cdot \R_{ij}}>^{irr}_{\alpha \beta \beta \alpha} = \frac{1}{10N}  <\sum_{ij}
\mur_{ij}^{dd}>^{irr}_{\alpha \beta \beta \alpha}
\end{equation}
where the last equality follows from the fact that irreducible kernels have a well-defined
value at $\kb=0$.

\section{Linear response for Smoluchowski dynamics}\label{lin}

The above developed formalism may also be applied to the calculation of the system
response
in the long-time regime, when the memory effects (caused by the relaxation of distribution
of particle positions) become important. As a first step towards solving this problem,
we apply the linear response theory to generalized
Smoluchowski equation, which governs the evolution of the particle distribution function
in
the configuration space, $P(\X,t)$ for the colloidal suspension. In the absence of
external
disturbances, the equilibrium distribution is given by
\begin{equation}
P_{eq} (\X)=e^{\here-\beta \phi(\X)}/Q,
\end{equation}
where $\phi$ is the potential of interparticle forces.

Next, we disturb the system by introducing the
imposed flow field $\vv_0(\rr)$ and external forces
$\E=(\eg_1,\dots,\eg_N)$  and calculate an induced mean force density and particle
current.
The evolution of  $P(\X,t)$ is then given by the
Generalized Smoluchowski Equation \cite{Pusey:1991}
\be
\pot P(\X,t) = {\boldsymbol{\cal D}(\X,t)} P(\X,t) \no \\
\ee
where the Smoluchowski operator, ${\cal D}(\X,t)$,
in the presence of the flow $\vv_0(\rr)$ and external forces
$\E$  reads
\be
 {\boldsymbol{\cal D}(\X,t)} \equiv \sum_{i,j=1}^{N} \frac{\partial}{
\partial \R_i}
\cdot \D_{ij}(\X) \cdot \left[ \frac{\partial}{\partial \R_j}
  + \beta (\boldsymbol{F}^{int}_i + \eg_i)  \right] + \frac{\partial}{\partial
\R_i}
\cdot \Ce^{\,t}_i(\X) \cdot \vv_0.
\ee
Here $\boldsymbol{D}(\X)$ is the diffusion matrix
\be
\boldsymbol{D}_{ij} = k_B T \mur^{tt}_{ij},
\ee
and
\be
\boldsymbol{F}^{int}_i = - \nabla_i \phi
\ee
are the interparticle forces.

For later use, we also introduce the adjoint Smoluchowski operator, $\Le$, which obeys
\be
\boldsymbol{\cal D} P_{eq}(\X)... = P_{eq}(\X) \boldsymbol{\cal L} ...
\label{wyciaganie}
\ee

Next, we find the mean particle current and force density. The former is
given by the following ensemble average
\begin{equation}
<\jj(\rr,\X)>_t \equiv <\sum_{i=1}^N \boldsymbol{\dot{R}}_i
\delta(\rr-\R_i)> = < \sum_{i=1}^N {\cal L} \R_i \delta(\rr-\R_i)>_t,
\end{equation}
where the symbol $< \ >_t$ denotes the average over $P(\X,t)$. Inserting the
explicit form of
adjoint Smoluchowski operator yields
 \begin{align}
<\jj(\rr,\X)>_t= < \sum_{i=1}^N \left\{ ( \beta^{-1} \pox  + \boldsymbol{
\cal F}^{int} + \E) \cdot \mur(\X) + \Ce(\X)
\vv_0 \right\}_i \delta(\rr-\R_i)>_t
\label{pp2}
\end{align}
where $\boldsymbol{\cal F}^{int} =
(\F_1^{int},\F_2^{int},...,\F_N^{int})$
and $\{ \ \}_i$ stands for i-th component (in particle indexes) of the
operator in brackets.
 For example
 \be
 \{ \E \cdot \mur(\X) \}_i = \sum_j {\eg}_j \cdot \mur_{ji} = \sum_j
 \mur_{ij} \cdot {\eg}_j.
\ee
where the symmetry of mobility matrix has been used in the last equality.
Moreover, in order to keep the notation simple,  from now on we denote translational part
of mobility matrix $\mur^{tt}$ simply by
 $\mur$, as only $\mur^{tt}$ appears in subsequent considerations. Analogous convention
 applies to $\C^{\,t}$ and $\Ce^{\,t}$, which will be written as
 $\C$ and $\Ce$ respectively.

By considerations similar to the above one can also find the mean force
density. As it has been
shown in \cite{Felderhof:1987} it is given by the formula
\be
<\f(\rr,\X)>_t = <(\beta^{-1} \pox + \boldsymbol{\cal F} + \E ) \cdot
\Ce(\X) - \zzr(\X) \vv_0>_t.
\label{sila}
\ee
In deriving the linear response formulas for the system of Brownian
particles the approach due to Felderhof and Jones
\cite{Felderhof-Jones:1983,Felderhof-Jones:1987:3} is
adopted. It is assumed that particles were at equilibrium in the infinite
past so that
$$
P(\X,t\rightarrow -\infty) = P_{eq} (\X)
$$
Subsequently the fields $\E$ and $\vv_0$ are turned on and the distribution
changes to
\begin{equation}
P(\X,t)=P_{eq}(\X) +\delta P(\X,t),
\label{ppp}
\end{equation}
\ii with $\delta P(\X,t)$ obeying (to the linear order in $\E$ and $\vv_0$):
\begin{equation}
\frac{\partial \delta P(\X,t)}{\partial t} - \boldsymbol{\cal D} \delta P =
-
\pox \cdot \left[ (\mur \E(t)+ \Ce \vv_0 (t) ) P_{eq} \right].
\end{equation}
\ii The solution of the above equation with initial
condition  $\delta P = 0$ for $t=-\infty$ is given by
\begin{equation}
\delta P(\X,t) = - P_{eq} \int_{-\infty}^t dt'
e^{\here\boldsymbol{\cal L}(t-t')} (\pox + \beta \boldsymbol{\cal F}) \cdot
[\mur \E (t')+ \Ce
\vv_0 (t')].
\label{dp}
\end{equation}
This allows us to rewrite the expressions for $<\f(\rr,\X)>_t$ and
$<\jj(\rr,\X)>_t$ as
 \begin{align}
&   <\jj(\rr)>_t  = <\jj>^{inst}_t + <\jj>^{ret}_t \equiv  \int \de \rr' (\Yje(\rr,\rr')
\eg(\rr',t) + \Yjv(\rr,
\rr') \vv_0(\rr',t))  +
\no \\
& \no \\
& + \int \de \rr' \int_{-\infty}^t \de t' (\Xje(\rr,\rr',t-t') \eg(\rr',t')
+ \Xjv(\rr,\rr',t-t')
\vv_0(\rr',t')), \no \\
& \label{wielkie3a} \\
&  <\f(\rr)>_t  =   <\f>^{inst}_t + <\f>^{ret}_t  \equiv \int \de \rr' (\Yfe(\rr,\rr') \eg
(
\rr',t) +
\Yfv(\rr,\rr') \vv_0(\rr',t)) + \no
\\
&  \no \\
&  \int \de \rr' \int_{-\infty}^t \de t' (\Xfe(\rr,\rr',t-t') \eg(\rr',t') +
\Xfv(\rr,\rr',t-t')
\vv_0(\rr',t')), \no \\
&
\label{wielkie3}
\end{align}
\ii where an auxiliary force field $\eg(\rr,t)$ was introduced, such that
\begin{equation}
\eg_{i}(t) = \int \delta(\rr - \R_i) \eg(\rr,t) \de \rr.
\end{equation}
\ii and we have singled out instantaneous and retarded part of system's
response (corresponding to averaging over $P_{eq}$ and $\delta P$ in Eq.~\eqref{ppp},
respectively). The former contribution appears
immediately
after $\eg$ or $\vv_0$ is turned on and follows the change of the external
perturbation, while the
latter describes memory effects due to the change of the distribution
function induced by
external forces.

Instantaneous response kernels introduced above are defined as follows
\begin{subequations}\label{ker2}
\begin{align}
 & \Yje(\rr,\rr') = <\sum_{i,j=1}^N \delta(\rr-\R_i) \mur_{ij} \delta(
\rr'-\R_j)>,  \label{ker2a} \\
&   \Yjv(\rr,\rr') = <\sum_{i=1}^N \delta(\rr-\R_i) \Ce_i(\rr') >, \label{
ker2b} \\
&   \Yfe(\rr,\rr') = <\sum_{j=1}^N \C(\rr)_j \delta(\rr'-\R_j)>, \label{
ker2c} \\
&   \Yfv(\rr,\rr') = <-\zzr(\rr,\rr')>,\label{ker2d}
\end{align}
\end{subequations}
\ii whereas time-dependent response kernels $\X$ are given by
\begin{subequations}\label{ker3}
\begin{align}
 & \Xje(\rr,\rr',t) = - \beta^{-1} <\sum_{i,j=1}^N \delta(\rr-\R_i) [\mur \cdot \labla]_i
 \evol{t} [(\rabla + \beta \boldsymbol{\cal F}) \cdot \mur^{tt} ]_j \delta(
\rr'-\R_j) >, \label{ker3a} \\
&   \Xjv(\rr,\rr',t) = - \beta^{-1}<\sum_{i=1}^N \delta(\rr-\R_i) [\mur
\cdot \labla]_i \evol{t}
(\rabla +\beta \boldsymbol{\cal F}) \cdot \Ce(\rr') >,  \label{ker3b} \\
&   \Xfe(\rr,\rr',t) = - \beta^{-1} <  \C(\rr) \cdot \labla \evol{t}
\sum_{j=1}^N [(\rabla +\beta
\boldsymbol{\cal F}) \cdot \mur ]_j \delta(\rr'-\R_j)>, \label{ker3c} \\
&   \Xfv(\rr,\rr',t) = - \beta^{-1} <\C(\rr) \cdot \labla \evol{t} (\rabla+
\beta \boldsymbol{\cal
F}) \cdot \Ce(\rr') >,
\label{ker3d}
\end{align}
\end{subequations}
where the symbols $\labla$ and $\rabla$ denote the operator
$\partial / \partial \X$ acting to the left and to the right respectively.

\section{The reduction of response kernels}

The instantaneous response kernels defined in (\ref{ker2a}-\ref{ker2d}) are reduced
according to the general formula \eqref{ared}.
Using the
scattering expansions \eqref{sf},\eqref{smu},\eqref{sc},\eqref{sce}  one
obtains then
\be
<\boldsymbol{Y}_{AB}> = <\boldsymbol{Y}_{Av}>^{irr} + <\boldsymbol{Y}_{Av}>^{irr} \gr <
\boldsymbol{Y}_{fB}>,
\label{fv}
\ee
where $A=j,E$ and $B=E,v$.

A somewhat harder task is to perform the reduction of the retarded response
kernels $\X$ given by Eq.~(\ref{ker3}). The general form of those kernels
is
$$\X =<A e^{\Le t} B>$$
with two operators $A$ and $B$ on both sides of the evolution operator
$e^{\Le t}$.
It is precisely the presence of the evolution operator in the kernels that makes the
reduction complicated. The procedure is as follows.
First,  the adjoint Smoluchowski operator
 \begin{equation}
\boldsymbol{\cal L} = [ \beta^{-1} \rabla + \boldsymbol{\cal F}] \cdot \mur
\cdot \rabla
\label{eo}
\end{equation}
\ii is decomposed as
 \begin{equation}
\boldsymbol{\cal L}(1,2,\dots,N)=\sum_{i,j=1}^N \Le_0(i)+ \delta \Le(1,2,
\dots,N),
\end{equation}
\ii where $\boldsymbol{\cal L}_0(i)$ is the single particle operator
 \begin{equation}
\Le_0(i) = D_0 \nabla_i^2,
\end{equation}
with $\nabla_i^2$ denoting the Laplacian with respect to $\R_i$ and $D_0$ - single
particle
diffusion coefficient.
It is worth noting that $\Le_0$ does not introduce any correlation between
the particles. Thus the evolution operator can be written as a series
 \begin{equation}
\evol{t} = S(t)+\int_0^t d\tau S(t-\tau) \delta L S(\tau) +
\int_0^t d\tau \int_0^{\tau} d\tau' S(t-\tau)
\delta L S(\tau-\tau') \delta L S(\tau) + ... ,
\label{ewo}
\end{equation}
\ii with
 \begin{equation}
S(1,2,\dots,N;t)=\prod_i^{N} S(i;t),
\end{equation}
\ii and
 \begin{equation}
S(i;t)=e^{\Le_0(i) t}.
\end{equation}
 Next,  the scattering expansions of the operators  $A$, $B$ and
$\delta L$ are performed.
 Then, after inserting the expansions into $<A \evol{t} B>$ one ends up with
the representation of the retarded response kernel as a sum of terms of the
following structure
\be
\begin{split} R_s(\SC,c)= & \int \mbox{d}1\dots\mbox{d}s \int_0^t d\tau
\int_0^{\tau_1}
\!\!  d\tau_2 \dots  \!\!\! \int_0^{
\tau_{n-1}} \!\!\!\! d\tau_n A'(t) S(t-\tau_1) \\  &
\delta L'(\tau_1) S(\tau_1-\tau_2) \delta L'(\tau_2) \dots S(\tau_n) B'(
\tau_n) c(1,\dots,s),
\label{rsc}
\end{split}
\ee
where $A'$, $B$ and $\delta L'$ stand for some elements of the scattering
expansions of $A$, $B$ and $\dl$ respectively and $s$ is the number of
particles appearing in the given term. The time variables have been
added to time-independent operators $\delta L'$, $A'$ and $B'$ just to
indicate their positions
relative to the evolution operators in the above integral.

\section{Diagrammatic expansion for time-dependent kernels}

Since the scattering expansion of time-dependent response kernels involves
more operators than the
instantaneous response kernels, we need to introduce new elements into the
diagrams, namely:

\begin{itemize}

\item the single-particle evolution operators $S(i,\tau-\tau')$ are
represented in the diagrams by horizontal solid lines (e-bonds):
$$
\begin{picture}(50,30)
\thicklines
\put(0,10){\line(1,0){40}}
\put(0,11){\line(1,0){40}}
\put(0,0){$\tau$}
\put(35,0){$\tau'$}
\thinlines
\end{picture}
$$

\item a dagger line \sztylet represents the two-body interparticle forces (
$\boldsymbol{\cal F}$ - bond)

\item single arrows ($\rightarrow$,$\leftarrow$) represent the operators
$\rabla$ and $\labla$ respectively

\item double arrows ($\Rightarrow$,$\Leftarrow$) represent
$\beta^{-1} \rabla$ and $\beta^{-1} \labla$ respectively

\end{itemize}

For example the diagram


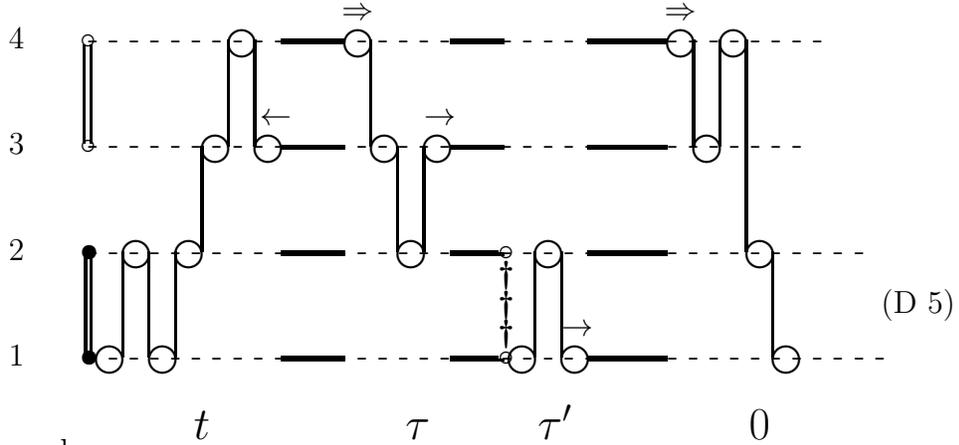
\hspace{3cm}
\begin{picture}(120,150)
\put(10,10){\kdp}
\put(20,50){\kd}
\put(30,10){\kd}
\put(40,50){\kd}
\put(50,90){\kd}
\put(60,130){\kdp}
\put(70,90){\kppl}
\thicklines
\put(13,11){\line(0,1){40}}
\put(23,51){\line(0,-1){40}}
\put(33,11){\line(0,1){40}}
\put(43,51){\line(0,1){40}}
\put(53,91){\line(0,1){40}}
\put(63,131){\line(0,-1){40}}
\put(73,10){\line(1,0){24}}
\put(73,50){\line(1,0){24}}
\put(73,90){\line(1,0){24}}
\put(73,130){\line(1,0){24}}
\put(73,11){\line(1,0){24}}
\put(73,51){\line(1,0){24}}
\put(73,91){\line(1,0){24}}
\put(73,131){\line(1,0){24}}
\put(40,-20){\Large $t$}
\put(120,-20){\Large $\tau$}
\put(170,-20){\Large $\tau'$}
\put(250,-20){\Large $0$}
\put(104,130){\klr}
\put(114,90){\kd}
\put(124,50){\kd}
\put(134,90){\kpr}
\put(107,131){\line(0,-1){40}}
\put(117,91){\line(0,-1){40}}
\put(127,91){\line(0,-1){40}}
\put(137,10){\line(1,0){20}}
\put(137,50){\line(1,0){20}}
\put(137,90){\line(1,0){20}}
\put(137,130){\line(1,0){20}}
\put(137,11){\line(1,0){18}}
\put(137,51){\line(1,0){18}}
\put(137,91){\line(1,0){20}}
\put(137,131){\line(1,0){20}}
\put(166,10){\kd}
\put(176,50){\kd}
\put(186,10){\kpr}
\put(169,11){\line(0,1){40}}
\put(179,51){\line(0,-1){40}}
\put(189,11){\line(1,0){30}}
\put(189,51){\line(1,0){30}}
\put(189,91){\line(1,0){30}}
\put(189,131){\line(1,0){30}}
\put(189,10){\line(1,0){30}}
\put(189,50){\line(1,0){30}}
\put(189,90){\line(1,0){30}}
\put(189,130){\line(1,0){30}}
\put(226,130){\klr}
\put(236,90){\kd}
\put(246,130){\kd}
\put(256,50){\kd}
\put(266,10){\kd}
\put(229,131){\line(0,-1){40}}
\put(239,91){\line(0,1){40}}
\put(249,131){\line(0,-1){80}}
\put(259,51){\line(0,-1){40}}
\put(-1,12){\line(0,1){38}}
\put(1,12){\line(0,1){37}}
\put(-1.5,92){\line(0,1){37}}
\put(1,92){\line(0,1){37}}
\put(0,48){- - - - - - - - - - - - - - - - - - - - - - - - - - - - - - - - -
- - - - - }
\put(0,8) {- - - - - - - - - - - - - - - - - - - - - - - - - - - - - - - - -
- - - - - -}
\put(0,88){- - - - - - - - - - - - - - - - - - - - - - - - - - - - - - - - -
- - }
\put(0,128){- - - - - - - - - - - - - - - - - - - - - - - - - - - - - - - -
- - - -}
\put(-30,48){2}
\put(-30,8){1}
\put(-30,88){3}
\put(-30,128){4}
\put(-3,48){$\boldsymbol{\bullet}$}
\put(-3,8){$\boldsymbol{\bullet}$}
\put(-3,88){$\circ$}
\put(-3,128){$\circ$}
\multiput(155,39)(0,-11){3}{$\boldsymbol{\dagger}$}
\put(155,48){$\circ$}
\put(155,8){$\circ$}
\put(300,28){(D 5)}
\end{picture}

%

\bigskip

\ii stands for the kernel

 \begin{align}
&  \int \de 1 \de 2 \ h(12) h(34) \Tilde{A}(1,2,3,4) \int_0^t d\tau \int_0^{
\tau} d\tau' \\
& S(1,2,3,4;t-\tau) \widetilde{\dl}_1(2,3,4)  S(1,2,3,4;\tau-\tau')
\widetilde{\dl}_2(1,2) S(1,2,3,4;\tau') \Tilde{B}(1,2,3,4), \no
\end{align}

\ii with the corresponding blocks given by

 \begin{align}
  \Tilde{A}(1,2,3,4)= & - \delta(\rr-\R_1) \M^<(1) \gr(12)
\M(2) \gr(21) \M(1) \no \\
&  \gr(12) \zr(2) \gr(23) \zr(3) \gr(34)
\z(4) \gr(43) \M^{>}(3) \labla_3, \no \\
  \widetilde{\dl}_1(2,3,4)= & \beta \rabla_4 \M^{<}(4) \gr(43) \M(3) \gr
(32) \M(2) \gr(23) \M^{>}(3) \rabla_3,  \no \\
  \widetilde{\dl}_2(1,2)= & \F(21) \M(1) \gr(12) \M(2) \gr(21) \M^{>}(1) \rabla_1,  \no \\
  \Tilde{B}(1,2,3,4)=& - \rabla_4 \M^{<}(4) \gr(43) \M(3) \gr(34)
\M(4) \gr(42) \no \\
&  \M(2) \gr(21) \M^>(1) \delta(\rr-\R_1). \no
\end{align}

The exact form of $\M$, $\M^{>}$ and $\M^{<}$ depends on the specific kernel to be
represented by the diagrams. For example, in the case of $\Xje$, we put
\be
\M(i)=\zr(i)
\ee
\be
\M^{<}(i)=\mo(i) \pe(i) \zo(i)
\ee
and
\be
\M^{>}(i)=  \zo(i) \pe(i) \mo(i)
\ee
As it is seen, the scattering sequences in time-dependent diagrams have a more complicated
structure than those encountered before, not only due to the presence of several
independent
blocks, but also due to the appearance of divergence operators $\labla$ and $\rabla$.

\section{Reduction of time-dependent diagrams}

The next task is to perform the reduction of the time-dependent diagrams
along similar lines to the approach presented previously - i.e. by
identification of connection lines.
The definition of a connection line is analogous to that  in
instantaneous response terms: an operator $\gr(i_k,i_{k+1})$  is called the
connection line of a term
$R_s(\SC,c)$ if the latter can be written as
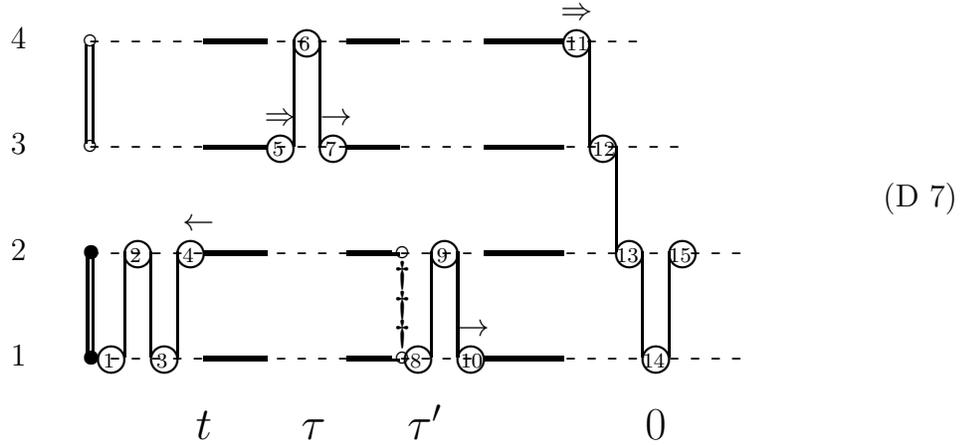
\begin{multline}
R_s(\SC,c) = \\
\int  \SC_1(i_1,i_2,\dots,i_k) c_1(i_1,i_2,\dots,i_k) \gr(i_k,i_{k+1}) \SC_2
(i_{k+1},\dots,i_s)
c_2(i_{k+1},\dots,i_s)
\mbox{d}1 \mbox{d}2 ... \mbox{d}s,
\label{redret}
\end{multline}
so that after the removal of $\gr(i_k,i_{k+1})$ the term $R_s(\SC,c)$
becomes a product of two
independent integrals. Integrals over time have been omitted in the above
expression as they are
irrelevant to our definition. The nodal line and nodal blocks for
terms $R_s(\SC,c)$ are also defined analogously to the instantaneous response
case. Thus, for example a diagram of the form

\hspace{3cm}
\begin{picture}(120,150)
\put(10,10){\kdp}
\put(20,50){\kd}
\put(30,10){\kd}
\put(40,50){\kppl}
\thicklines
\put(13,11){\line(0,1){40}}
\put(23,51){\line(0,-1){40}}
\put(33,11){\line(0,1){40}}
\put(43,10){\line(1,0){24}}
\put(43,50){\line(1,0){24}}
\put(43,90){\line(1,0){24}}
\put(43,130){\line(1,0){24}}
\put(43,11){\line(1,0){24}}
\put(43,51){\line(1,0){24}}
\put(43,91){\line(1,0){24}}
\put(43,131){\line(1,0){24}}
\put(40,-20){\Large $t$}
\put(80,-20){\Large $\tau$}
\put(120,-20){\Large $\tau'$}
\put(210,-20){\Large $0$}
\put(74,90){\klr}
\put(84,130){\kd}
\put(94,90){\kpr}
\put(77,91){\line(0,1){40}}
\put(87,91){\line(0,1){40}}
\put(97,10){\line(1,0){20}}
\put(97,50){\line(1,0){20}}
\put(97,90){\line(1,0){20}}
\put(97,130){\line(1,0){20}}
\put(97,11){\line(1,0){17}}
\put(97,51){\line(1,0){17}}
\put(97,91){\line(1,0){20}}
\put(97,131){\line(1,0){20}}
\put(126,10){\kd}
\put(136,50){\kd}
\put(146,10){\kpr}
\put(129,11){\line(0,1){40}}
\put(139,51){\line(0,-1){40}}
\put(149,11){\line(1,0){30}}
\put(149,51){\line(1,0){30}}
\put(149,91){\line(1,0){30}}
\put(149,131){\line(1,0){30}}
\put(149,10){\line(1,0){30}}
\put(149,50){\line(1,0){30}}
\put(149,90){\line(1,0){30}}
\put(149,130){\line(1,0){30}}
\put(186,130){\klr}
\put(196,90){\kd}
\put(206,50){\kd}
\put(216,10){\kd}
\put(226,50){\kd}
\put(189,131){\line(0,-1){40}}
\put(199,91){\line(0,-1){40}}
\put(209,51){\line(0,-1){40}}
\put(219,51){\line(0,-1){40}}
\put(-1,12){\line(0,1){38}}
\put(1,12){\line(0,1){37}}
\put(-1.5,92){\line(0,1){37}}
\put(1,92){\line(0,1){37}}
\put(0,48){- - - - - - - - - - - - - - - - - - - - - - - - - - - - - - - - -
 }
\put(0,8) {- - - - - - - - - - - - - - - - - - - - - - - - - - - - - - - -}
\put(0,88){- - - - - - - - - - - - - - - - - - - - - - - - - - - - -}
\put(0,128){- - - - - - - - - - - - - - - - - - - - - - - - - - -}
\put(-30,48){2}
\put(-30,8){1}
\put(-30,88){3}
\put(-30,128){4}
\put(-3,48){$\boldsymbol{\bullet}$}
\put(-3,8){$\boldsymbol{\bullet}$}
\put(-3,88){$\circ$}
\put(-3,128){$\circ$}
\multiput(115,39)(0,-11){3}{$\boldsymbol{\dagger}$}
\put(115,48){$\circ$}
\put(115,8){$\circ$}
\put(300,68){(D 6)}
\end{picture}

\vspace{1cm}

has a single connection line (the one joining the particles $2$ and $3$).
This is also a nodal line of this diagram.

However, because of the fact that retarded response terms consist of a
number of individual operators $A'$,
$\delta L'(\tau_1)$, $\delta L'(\tau_2)\dots$ the nodal structure of
$R_s(\SC,c)$ is usually very
complicated and in general it is impossible to apply the concept of block
distribution function here. To analyse the nodal structure of
retarded response kernels an ordering of the graph nodes is introduced first.
Namely, moving along the graph from the left to the right we index all the
nodes with the subsequent natural numbers. For the diagram (D 6) one gets

\hspace{3cm}
\begin{picture}(120,150)
\put(10,10){\kdp}
\put(20,50){\kd}
\put(30,10){\kd}
\put(40,50){\kppl}
\put(5,9){${}_{1}$}
\put(15,49){${}_{2}$}
\put(25,9){${}_{3}$}
\put(35,49){${}_{4}$}
\thicklines
\put(13,11){\line(0,1){40}}
\put(23,51){\line(0,-1){40}}
\put(33,11){\line(0,1){40}}
\put(43,10){\line(1,0){24}}
\put(43,50){\line(1,0){24}}
\put(43,90){\line(1,0){24}}
\put(43,130){\line(1,0){24}}
\put(43,11){\line(1,0){24}}
\put(43,51){\line(1,0){24}}
\put(43,91){\line(1,0){24}}
\put(43,131){\line(1,0){24}}
\put(40,-20){\Large $t$}
\put(80,-20){\Large $\tau$}
\put(120,-20){\Large $\tau'$}
\put(210,-20){\Large $0$}
\put(74,90){\klr}
\put(84,130){\kd}
\put(94,90){\kpr}
\put(69,89){${}_{5}$}
\put(79,129){${}_{6}$}
\put(89,89){${}_{7}$}
\put(77,91){\line(0,1){40}}
\put(87,91){\line(0,1){40}}
\put(97,10){\line(1,0){20}}
\put(97,50){\line(1,0){20}}
\put(97,90){\line(1,0){20}}
\put(97,130){\line(1,0){20}}
\put(97,11){\line(1,0){17}}
\put(97,51){\line(1,0){17}}
\put(97,91){\line(1,0){20}}
\put(97,131){\line(1,0){20}}
\put(126,10){\kd}
\put(136,50){\kd}
\put(146,10){\kpr}
\put(121,9){${}_{8}$}
\put(131,49){${}_{9}$}
\put(140,9){${}_{10}$}
\put(129,11){\line(0,1){40}}
\put(139,51){\line(0,-1){40}}
\put(149,11){\line(1,0){30}}
\put(149,51){\line(1,0){30}}
\put(149,91){\line(1,0){30}}
\put(149,131){\line(1,0){30}}
\put(149,10){\line(1,0){30}}
\put(149,50){\line(1,0){30}}
\put(149,90){\line(1,0){30}}
\put(149,130){\line(1,0){30}}
\put(186,130){\klr}
\put(196,90){\kd}
\put(206,50){\kd}
\put(216,10){\kd}
\put(226,50){\kd}
\put(180,129){${}_{11}$}
\put(190,89){${}_{12}$}
\put(199,49){${}_{ 13}$}
\put(209,9){${}_{ 14}$}
\put(219,49){${}_{15}$}
\put(189,131){\line(0,-1){40}}
\put(199,91){\line(0,-1){40}}
\put(209,51){\line(0,-1){40}}
\put(219,51){\line(0,-1){40}}
\put(-1,12){\line(0,1){38}}
\put(1,12){\line(0,1){37}}
\put(-1.5,92){\line(0,1){37}}
\put(1,92){\line(0,1){37}}
\put(0,48){- - - - - - - - - - - - - - - - - - - - - - - - - - \ \ - \ \ - -
-
 }
\put(0,8) {- - - - - - - - - - - - - - - - - - - - - - - - - - -
 \ \ - - - -}
\put(0,88){- - - - - - - - - - - - - - - - - - - - - - - - - - - - -}
\put(0,128){- - - - - - - - - - - - - - - - - - - - - - - - - - -}
\put(-30,48){2}
\put(-30,8){1}
\put(-30,88){3}
\put(-30,128){4}
\put(-3,48){$\boldsymbol{\bullet}$}
\put(-3,8){$\boldsymbol{\bullet}$}
\put(-3,88){$\circ$}
\put(-3,128){$\circ$}
\multiput(115,39)(0,-11){3}{$\boldsymbol{\dagger}$}
\put(115,48){$\circ$}
\put(115,8){$\circ$}
\put(300,68){(D 7)}
\end{picture}

\vspace{1cm}

The above defined ordering allows us to introduce the notion of a {\bf
proper diagram}.
To define it, let us consider a diagram $R_s$ with a scattering structure
$\SC(i_1,i_2,\dots,i_s)$ and a nodal line $\gr(i_k,i_{k+1})$ such that
\be
\SC(i_1,i_2,\dots,i_s) = \SC_1(i_1,i_2,\dots,i_k) \gr(i_k,i_{k+1}) \SC_2(i_{
k+1},\dots,i_s).
\ee
The diagram $R_s$ will be called proper if
all the operators in which the particles from $\{i_1,i_2,\dots,i_k\}$
appear have smaller indexes than these in which
$\{i_{k+1},\dots,i_s\}$ appear. Thus the diagram (D 2) is not proper whereas
the one of the form

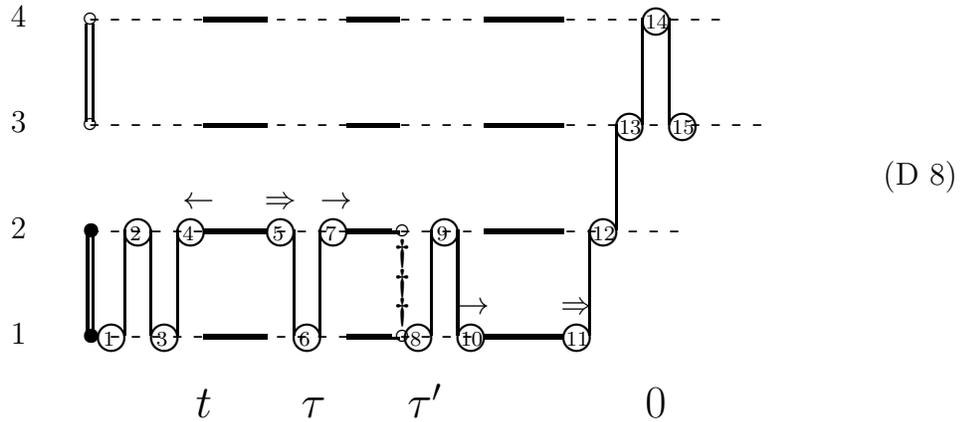
\hspace{3cm}
\begin{picture}(120,150)
\put(10,10){\kdp}
\put(20,50){\kd}
\put(30,10){\kd}
\put(40,50){\kppl}
\put(5,9){${}_{1}$}
\put(15,49){${}_{2}$}
\put(25,9){${}_{3}$}
\put(35,49){${}_{4}$}
\thicklines
\put(13,11){\line(0,1){40}}
\put(23,51){\line(0,-1){40}}
\put(33,11){\line(0,1){40}}
\put(43,10){\line(1,0){24}}
\put(43,50){\line(1,0){24}}
\put(43,90){\line(1,0){24}}
\put(43,130){\line(1,0){24}}
\put(43,11){\line(1,0){24}}
\put(43,51){\line(1,0){24}}
\put(43,91){\line(1,0){24}}
\put(43,131){\line(1,0){24}}
\put(40,-20){\Large $t$}
\put(80,-20){\Large $\tau$}
\put(120,-20){\Large $\tau'$}
\put(210,-20){\Large $0$}
\put(74,50){\klr}
\put(84,10){\kd}
\put(94,50){\kpr}
\put(69,49){${}_{5}$}
\put(79,9){${}_{6}$}
\put(89,49){${}_{7}$}
\put(77,11){\line(0,1){40}}
\put(87,11){\line(0,1){40}}
\put(97,10){\line(1,0){20}}
\put(97,50){\line(1,0){20}}
\put(97,90){\line(1,0){20}}
\put(97,130){\line(1,0){20}}
\put(97,11){\line(1,0){17}}
\put(97,51){\line(1,0){17}}
\put(97,91){\line(1,0){20}}
\put(97,131){\line(1,0){20}}
\put(126,10){\kd}
\put(136,50){\kd}
\put(146,10){\kpr}
\put(121,9){${}_{8}$}
\put(131,49){${}_{9}$}
\put(140,9){${}_{10}$}
\put(129,11){\line(0,1){40}}
\put(139,51){\line(0,-1){40}}
\put(149,11){\line(1,0){30}}
\put(149,51){\line(1,0){30}}
\put(149,91){\line(1,0){30}}
\put(149,131){\line(1,0){30}}
\put(149,10){\line(1,0){30}}
\put(149,50){\line(1,0){30}}
\put(149,90){\line(1,0){30}}
\put(149,130){\line(1,0){30}}
\put(186,10){\klr}
\put(196,50){\kd}
\put(206,90){\kd}
\put(216,130){\kd}
\put(226,90){\kd}
\put(180,9){${}_{11}$}
\put(190,49){${}_{12}$}
\put(200,89){${}_{13}$}
\put(210,129){${}_{14}$}
\put(220,89){${}_{15}$}
\put(189,11){\line(0,1){40}}
\put(199,51){\line(0,1){40}}
\put(209,91){\line(0,1){40}}
\put(219,131){\line(0,-1){40}}
\put(-1,12){\line(0,1){38}}
\put(1,12){\line(0,1){37}}
\put(-1.5,92){\line(0,1){37}}
\put(1,92){\line(0,1){37}}
\put(0,48){- - - - - - - - - - - - - - - - - - - - - - - - - \ \ - - -}
\put(0,8) {- - - - - - - - - - - - - - - - - - - - - - -}
\put(0,88){- - - - - - - - - - - - - - - - - - - - - - - - - - \ \ - \ \ - - -
- }
\put(0,128){- - - - - - - - - - - - - - - - - - - - - - - - - - - \ \ - - -}
\put(-30,48){2}
\put(-30,8){1}
\put(-30,88){3}
\put(-30,128){4}
\put(-3,48){$\boldsymbol{\bullet}$}
\put(-3,8){$\boldsymbol{\bullet}$}
\put(-3,88){$\circ$}
\put(-3,128){$\circ$}
\multiput(115,39)(0,-11){3}{$\boldsymbol{\dagger}$}
\put(115,48){$\circ$}
\put(115,8){$\circ$}
\put(300,68){(D 8)}
\end{picture}

\vspace{1cm}

is proper.  Note that the
definition of a proper term
concerns only the scattering structure in a diagram, the correlation
structure is irrelevant here.

\subsection{Nodal structure of time dependent kernels}

As the evolution diagrams consist of many different building blocks ($A$,
$B$ and $\dl$) no wonder that their nodal structure is much more complicated
than that of the instantaneous response diagrams analyzed in Section \ref{redi}.
For example the diagram

\hspace{3cm}
\begin{picture}(120,220)
\put(10,10){\kdp}
\put(20,50){\kd}
\put(30,10){\kd}
\put(40,130){\kd}
\put(50,90){\kppl}
\thicklines
\put(13,11){\line(0,1){40}}
\put(23,51){\line(0,-1){40}}
\put(33,11){\line(0,1){120}}
\put(43,131){\line(0,-1){40}}
\put(53,90){\line(1,0){24}}
\put(53,130){\line(1,0){24}}
\put(53,91){\line(1,0){24}}
\put(53,131){\line(1,0){24}}
\put(40,-50){\Large $t$}
\put(90,-50){\Large $\tau$}
\put(140,-50){\Large $\tau'$}
\put(210,-50){\Large $0$}
\put(84,170){\klr}
\put(94,90){\kd}
\put(104,130){\kpr}
\put(87,91){\line(0,1){80}}
\put(97,91){\line(0,1){40}}
\put(107,90){\line(1,0){20}}
\put(107,130){\line(1,0){20}}
\put(107,91){\line(1,0){20}}
\put(107,131){\line(1,0){20}}
\put(136,90){\klr}
\put(146,130){\kd}
\put(156,-30){\kpr}
\put(139,131){\line(0,-1){40}}
\put(149,131){\line(0,-1){160}}
\put(159,91){\line(1,0){30}}
\put(159,131){\line(1,0){30}}
\put(159,90){\line(1,0){30}}
\put(159,130){\line(1,0){30}}
\put(196,90){\klr}
\put(206,130){\kd}
\put(216,210){\kd}
\put(199,91){\line(0,1){40}}
\put(209,131){\line(0,1){80}}
\put(-1,12){\line(0,1){38}}
\put(1,12){\line(0,1){37}}
\put(-1.5,92){\line(0,1){37}}
\put(1,92){\line(0,1){37}}
\put(0,48){- - -}
\put(0,8) {- - - - - }
\put(0,88){- - - - - - - - - - - - - - - - - - - - - - - - - - - - - - - - -
- }
\put(0,128){- - - - - - - - - - - - - - - - - - - - - - - - - - - - - - -}
\put(-30,48){3}
\put(-30,8){2}
\put(-30,88){4}
\put(-30,128){5}
\put(-30,168){6}
\put(-30,208){7}
\put(-30,-32){1}
\put(-3,48){$\boldsymbol{\bullet}$}
\put(-3,8){$\boldsymbol{\bullet}$}
\put(-3,88){$\circ$}
\put(-3,128){$\circ$}
\put(300,68){(D 9)}
\end{picture}

\vspace{3cm}

has the nodal structure of the form


\begin{picture}(120,220)
\thicklines
\put(50,11){\line(1,0){40}}
\put(50,11){\line(0,1){40}}
\put(90,11){\line(0,1){120}}
\put(50,51){\line(1,0){40}}

\put(90,131){\line(1,0){200}}
\put(90,91){\line(1,0){200}}

\put(290,91){\line(0,1){40}}

\put(180,107){4,5}
\put(60,26){2,3}
\put(240,10){1}
\put(160,175){6}
\put(300,187){7}

\put(220,91){\line(0,-1){100}}
\put(220,-9){\line(1,0){40}}
\put(220,31){\line(1,0){40}}
\put(260,-9){\line(0,1){40}}


\put(290,131){\line(0,1){80}}
\put(290,211){\line(1,0){40}}
\put(290,171){\line(1,0){40}}
\put(330,171){\line(0,1){40}}

\put(140,161){\line(0,1){40}}
\put(140,201){\line(1,0){40}}
\put(140,161){\line(1,0){40}}
\put(180,131){\line(0,1){70}}



\end{picture}

\vspace{1cm}

In graph theory the above structure is called {\bf a tree}: a connected
graph which do not contain
any circuits (the lack of circuits stems directly from the definition of the
nodal line). Unfortunately, the presence of many branches makes it
impossible to apply in this case the methods developed in Section
 \ref{redi}. In particular, the block distribution function cannot be defined
on the nodal structure like that of the diagram (D 4), as it lacks the linear
ordering.

Luckily,
 the nodal structure graph of the proper terms is simpler. Namely, in the proper diagrams
by the definition left-right ordering of the vertices is compatible with the nodal
structure. Hence the nodal graph of a proper diagram is a
 {\bf simple chain} - a tree with two terminal vertices only.

For
example, the proper diagram of the form

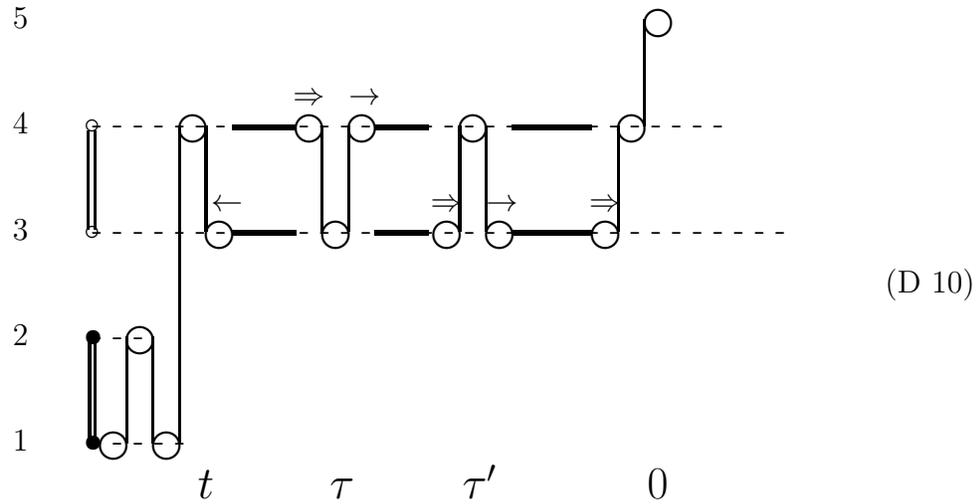
\hspace{3cm}
\begin{picture}(120,200)
\put(10,10){\kdp}
\put(20,50){\kd}
\put(30,10){\kd}
\put(40,130){\kd}
\put(50,90){\kppl}
\thicklines
\put(13,11){\line(0,1){40}}
\put(23,51){\line(0,-1){40}}
\put(33,11){\line(0,1){120}}
\put(43,131){\line(0,-1){40}}
\put(53,90){\line(1,0){24}}
\put(53,130){\line(1,0){24}}
\put(53,91){\line(1,0){24}}
\put(53,131){\line(1,0){24}}
\put(40,-10){\Large $t$}
\put(90,-10){\Large $\tau$}
\put(140,-10){\Large $\tau'$}
\put(210,-10){\Large $0$}
\put(84,130){\klr}
\put(94,90){\kd}
\put(104,130){\kpr}
\put(87,91){\line(0,1){40}}
\put(97,91){\line(0,1){40}}
\put(107,90){\line(1,0){20}}
\put(107,130){\line(1,0){20}}
\put(107,91){\line(1,0){20}}
\put(107,131){\line(1,0){20}}
\put(136,90){\klr}
\put(146,130){\kd}
\put(156,90){\kpr}
\put(139,131){\line(0,-1){40}}
\put(149,131){\line(0,-1){40}}
\put(159,91){\line(1,0){30}}
\put(159,131){\line(1,0){30}}
\put(159,90){\line(1,0){30}}
\put(159,130){\line(1,0){30}}
\put(196,90){\klr}
\put(206,130){\kd}
\put(216,170){\kd}
\put(199,91){\line(0,1){40}}
\put(209,131){\line(0,1){40}}
\put(-1,12){\line(0,1){38}}
\put(1,12){\line(0,1){37}}
\put(-1.5,92){\line(0,1){37}}
\put(1,92){\line(0,1){37}}
\put(0,48){- - -}
\put(0,8) {- - - - - }
\put(0,88){- - - - - - - - - - - - - - - - - - - - - - - - - - - - - - - - -
- }
\put(0,128){- - - - - - - - - - - - - - - - - - - - - - - - - - - - - - -}
\put(-30,48){2}
\put(-30,8){1}
\put(-30,88){3}
\put(-30,128){4}
\put(-30,168){5}
\put(-3,48){$\boldsymbol{\bullet}$}
\put(-3,8){$\boldsymbol{\bullet}$}
\put(-3,88){$\circ$}
\put(-3,128){$\circ$}
\put(300,68){(D 10)}
\end{picture}

\vspace{1cm}

has a chain-like nodal structure graph of the form

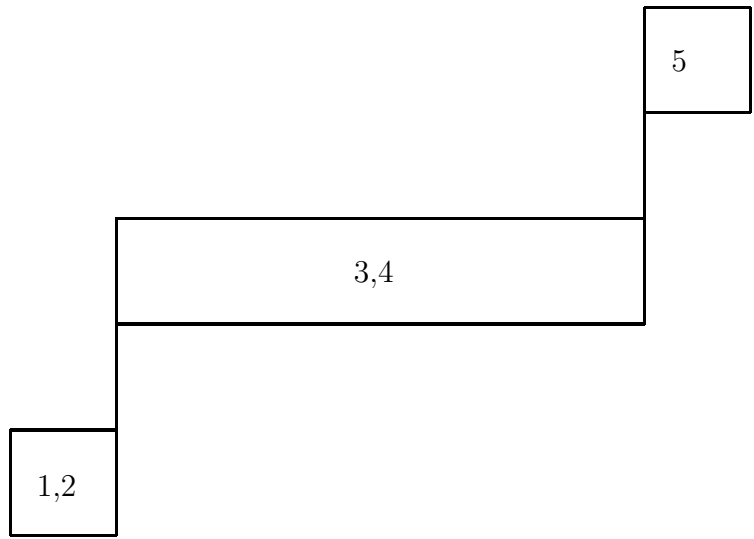
\begin{picture}(120,225)
\thicklines
\put(50,11){\line(1,0){40}}
\put(50,11){\line(0,1){40}}
\put(90,11){\line(0,1){120}}

\put(50,51){\line(1,0){40}}

\put(90,131){\line(1,0){200}}
\put(90,91){\line(1,0){200}}

\put(290,91){\line(0,1){40}}

\put(180,107){3,4}
\put(60,26){1,2}
\put(300,187){5}

\put(290,131){\line(0,1){80}}
\put(290,211){\line(1,0){40}}
\put(290,171){\line(1,0){40}}
\put(330,171){\line(0,1){40}}

\end{picture}

\bigskip

Thus in proper diagrams, nodal lines divide the particles $i_1,\dots,i_s$ into
nodal blocks $C_1,C_2,\dots$ which can be ordered according to the place in the chain.
This means that the nodal structure can again be written in the form
$C_1|C_2|...|C_k$, where $C_1$, $C_2$, ....,$C_k$ come one after another in
the time integral \eqref{rsc}. For such a structure  a block
distribution function can again be defined by Eq.~\eqref{bb}.

As it was mentioned, these concepts cannot be applied in the case of
improper terms. However, it may be
shown \cite{Szymczak:2001} that in the
thermodynamic limit the
sum of all improper diagrams in the expansion of a given time-dependent kernel vanishes.
 (We give see  the sketch of the proof in the Appendix.)
Therefore in the subsequent analysis we can consider proper terms
only. The fact that the time-dependent diagrams have a chain-like structure is an
important
result, since it allows us to use a concept of block distribution function and carry out
the reduction procedure in the case of time-dependent response. This element was missed by
the authors of Ref. \cite{Felderhof-Jones:1987:3} who applied directly the block-
distribution function analysis in their studies on linear response theory of viscosity,
without showing first that the structure of the terms in respective scattering
expansion is indeed chain-like.

Because of the chain-like form of the diagrams, it is now relatively easy to sum the
proper terms which share a similar nodal structure. For example, the
proper diagrams of the kernel $$\X =<A e^{\Le t} B>$$ may be divided in the
following groups

\begin{enumerate}
\item Diagrams with the articulation line in A-block
\item Diagrams with the articulation line in $\delta L$ -block
\item Diagrams with the articulation line in B-block
\item Irreducible diagrams.
\end{enumerate}

Below, the reduction procedure is carried out for the diagrams of each type

\begin{enumerate}
\item Proper diagrams with the articulation line inside
A-block are of the form

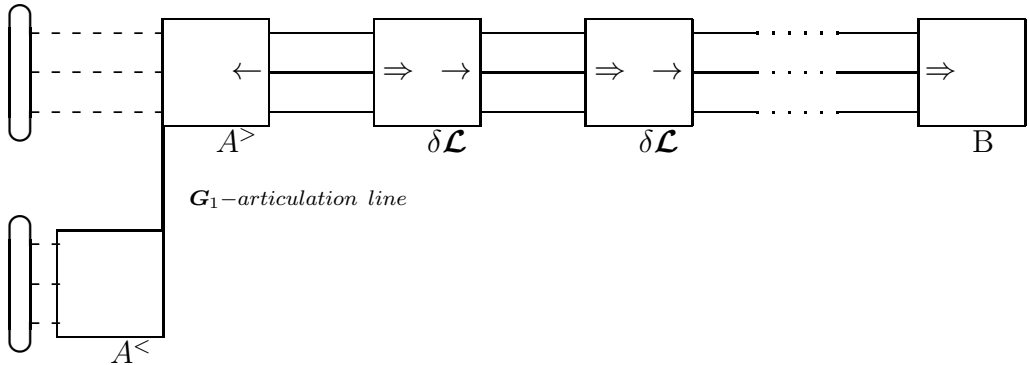
\begin{picture}(100,180)
\put(0,40){\framebox(40,40){{$\scriptstyle {}$ }}}
\put(40,120){\framebox(40,40){{$ \ \ \ \ \ \ \ \leftarrow $ }}}
\put(120,120){\framebox(40,40){{$ \Rightarrow \ \, \ \rightarrow $}}}
\put(200,120){\framebox(40,40){{$ \Rightarrow \ \, \ \rightarrow $}}}
\put(326,120){\framebox(40,40){{$ \Rightarrow \ \ \ \ \ \ $}}}
\thicklines
\put(40,80){\line(0,1){40}}
\thinlines
\put(50,90){{$\scriptstyle \gr_1- articulation \ line$}}
\put(20,30){$A^{<}$}
\put(60,110){$A^{>}$}
\put(140,110){$\dl $}
\put(220,110){$\dl $}
\put(346,110){B}
\put(-22,77){\kore}
\put(-22,157){\kore}
\put(80,125){\line(1,0){40}}
\put(80,140){\line(1,0){40}}
\put(80,155){\line(1,0){40}}
\put(-10,42.5){- - }
\put(-10,57.5){- - }
\put(-10,72.5){- - }
\put(-10,122.5){- - - - - - -}
\put(-10,137.5){- - - - - - -}
\put(-10,152.5){- - - - - - - }
\put(160,125){\line(1,0){40}}
\put(160,140){\line(1,0){40}}
\put(160,155){\line(1,0){40}}
\put(240,125){\line(1,0){25}}
\put(296,125){\line(1,0){30}}
\put(240,155){\line(1,0){25}}
\put(296,155){\line(1,0){30}}
\put(240,140){\line(1,0){25}}
\put(296,140){\line(1,0){30}}
\thicklines
\qbezier[5](265,155)(280,155)(295,155)
\qbezier[5](265,125)(280,125)(295,125)
\qbezier[5](265,140)(280,140)(295,140)
\thinlines
\end{picture}

\ii where the ovals stand for correlation functions and the divergence operators in each
block are marked

The kernel $A$ may be now reduced analogously to \eqref{ared} which gives
\be
\X_1=<A^{<}>^{irr} \gr <A^> e^{\Le t} B>
\label{pierw}
\ee

\item The proper diagrams with the articulation line inside
$\dl$-block are of the form

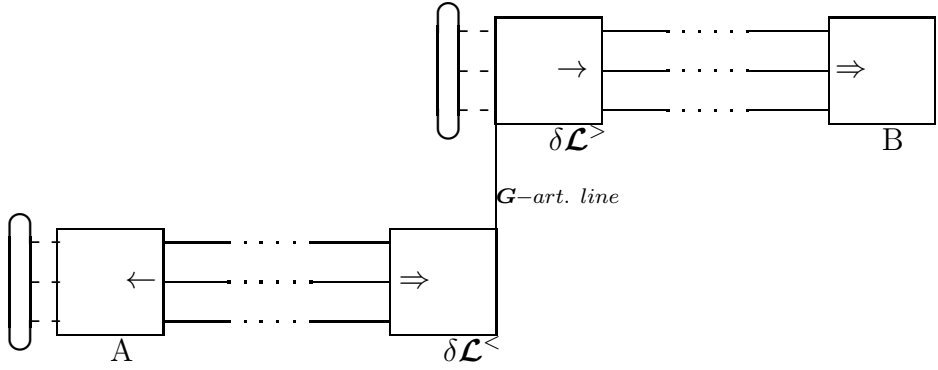
\begin{picture}(100,160)
\put(0,20){\framebox(40,40){{$\ \ \ \ \ \ \ \leftarrow $ }}}
\put(126,20){\framebox(40,40){{$ \Rightarrow \ \, \ \ \ \ $}}}

\put(166,100){\framebox(40,40){{$ \ \ \ \, \ \rightarrow $}}}
\put(292,100){\framebox(40,40){{$\Rightarrow \ \ \ \ \ \ $}}}

\put(166,60){\line(0,1){60}}
\put(166,70){$\scriptstyle \gr - art. \ line $}
\put(-22,57){\kore}
\put(152,102.5){- -}
\put(152,117.5){- -}
\put(152,132.5){- -}
\put(-10,22.5){- - }
\put(-10,37.5){- - }
\put(-10,52.5){- - }
\put(140,137){\kore}

\put(20,10){A}
\put(146,10){$\dl^{<}$}
\put(186,90){$\dl^{>}$}
\put(312,90){B}
\put(206,105){\line(1,0){25}}
\put(262,105){\line(1,0){30}}
\put(206,135){\line(1,0){25}}
\put(262,135){\line(1,0){30}}
\put(206,120){\line(1,0){25}}
\put(262,120){\line(1,0){30}}
\thicklines
\qbezier[5](231,135)(246,135)(261,135)
\qbezier[5](231,105)(246,105)(261,105)
\qbezier[5](231,120)(246,120)(261,120)
\thinlines
\put(40,25){\line(1,0){25}}
\put(96,25){\line(1,0){30}}
\put(40,55){\line(1,0){25}}
\put(96,55){\line(1,0){30}}
\put(40,40){\line(1,0){25}}
\put(96,40){\line(1,0){30}}
\thicklines
\qbezier[5](65,55)(80,55)(96,55)
\qbezier[5](65,25)(80,25)(96,25)
\qbezier[5](65,40)(80,40)(96,40)
\thinlines
\end{picture}

\bigskip

Thus, after the reduction, the diagrams of $\X_2$ sum up to

\begin{equation}
X_2=\int_0^{t} \de \tau <A e^{\Le (t-\tau)} \delta L^{<}>^{irr} \gr <\delta
L^{>} e^{\Le \tau} B>.
\end{equation}

\item Nonvanishing diagrams with articulation line inside
B-block are of the form

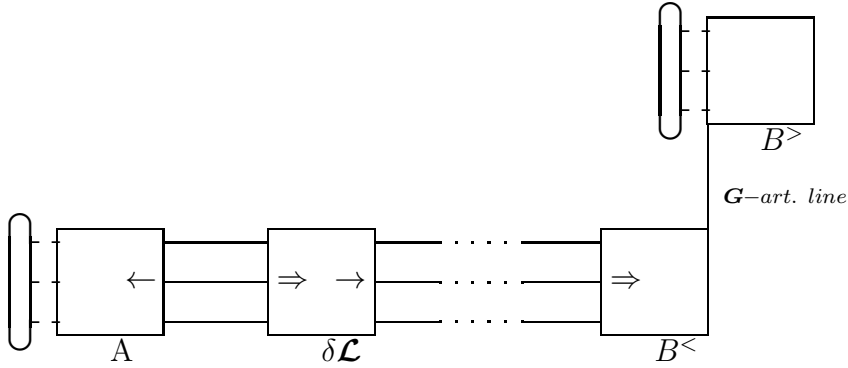
\begin{picture}(100,160)
\put(0,20){\framebox(40,40){{$\ \ \ \ \ \ \ \leftarrow $ }}}
\put(80,20){\framebox(40,40){{$ \Rightarrow \ \, \ \rightarrow $}}}
\put(206,20){\framebox(40,40){{$ \Rightarrow \ \, \ \ \ \ $}}}
\put(246,100){\framebox(40,40){{$ {} $}}}
\put(246,60){\line(0,1){40}}
\put(252,70){$\scriptstyle \gr - art. \ line $}
\put(224,137){\kore}
\put(236,102.5){- - }
\put(236,117.5){- - }
\put(236,132.5){- - }
\put(-22,57){\kore}
\put(-10,22.5){- - }
\put(-10,37.5){- - }
\put(-10,52.5){- - }
\put(20,10){A}
\put(100,10){$\dl $}
\put(226,10){$B^{<}$}
\put(266,90){$B^{>}$}
\put(40,25){\line(1,0){40}}
\put(40,40){\line(1,0){40}}
\put(40,55){\line(1,0){40}}

\put(120,25){\line(1,0){25}}
\put(176,25){\line(1,0){30}}
\put(120,55){\line(1,0){25}}
\put(176,55){\line(1,0){30}}
\put(120,40){\line(1,0){25}}
\put(176,40){\line(1,0){30}}
\thicklines
\qbezier[5](145,55)(160,55)(176,55)
\qbezier[5](145,25)(160,25)(176,25)
\qbezier[5](145,40)(160,40)(176,40)
\thinlines
\end{picture}
\bigskip

and they sum up to
\begin{equation}
X_3= <A e^{\Le t} \delta B^>>^{irr} \gr <B^{>}>.
\end{equation}
\item Finally, the irreducible diagrams give
 \begin{equation}
\X_4=<A e^{\Le t} \delta B >^{irr}.
\label{osta}
\end{equation}

\end{enumerate}

Eventually, summing up (\ref{pierw}-\ref{osta}) we get for the kernel $\X$
the following expression
 \begin{equation}
\begin{split}
& \X(t) = <A e^{\Le t} \delta B >= <A e^{\Le t} \delta B >^{irr} +
<A^{<}>^{irr} \gr <A^> e^{\Le t} B> + \\
& \int_0^{t} \de \tau <A e^{\Le (t-\tau)} {\dl}^{<}>^{irr} \gr <{\dl}^{>} e^{\Le \tau} B>
+
<A e^{\Le t} B^<>^{irr} \gr <B^{>}>
\end{split}
\end{equation}
The above algorithm may be now used to reduce the time-dependent kernels
defined in Eq.~\eqref{ker3}. Namely, the analysis of the scattering structures of both
retarded and instantaneous kernels (\ref{ker2a}-\ref{ker3d}) leads to
\begin{equation}
\begin{split}
& \boldsymbol{X}_{AB}(t) = \boldsymbol{X}_{AB}^{irr}(t)+\boldsymbol{Y}_{Av}^{irr} \gr
\boldsymbol{X}_{fB}(t) + \int_0^{t} \de \tau \boldsymbol{X}_{Av}^
{irr}(t-\tau) \gr \boldsymbol{X}_{fB}
(\tau) + \boldsymbol{X}_{Av}^{irr}(t) \gr \boldsymbol{Y}_{fB}^{irr}.
\end{split}
\label{ixi}
\end{equation}
where again $A=j,E$ and $B=E,v$.

\section{Effective equations}

The reductions of instantaneous kernels $\Y$  and time-dependent kernels
$\X$ carried out above may now be used to obtain the effective equations
governing the dynamics of suspensions. Namely, using \eqref{fv} in
Eq.~\eqref{wielkie3a} we get the following expression for the instantaneous
part of the current
\begin{equation}
\begin{split}
<\jj>_t^{ins} = \Yje \eg + \Yjv \vv_0 =
\Yje^{irr} \eg + \Yjv^{irr} \vv_0
 + \Yjv^{irr} \gr <\f>^{ins}
\end{split}
\end{equation}
where the definition of the instantaneous force density \eqref{wielkie3} was
used.

The retarded part of the current may be similarly obtained from the
reduction formulae \eqref{ixi}.
\begin{equation}
\begin{split}
& <\jj>^{ret}_t  =  \int_{-\infty}^t \de t' (\Xje^{irr}(t-t') \eg(t') + \Xjv
^{irr}(t-t') \vv_0(t'))
+ \\
& +  \Yjv^{irr} \gr \int_{-\infty}^t \de t' \Bigl( \Xfe(t-t') \eg(t') + \Xfv
(t-t') \vv_0(t')
\Bigr) + \\ & +
\int_{-\infty}^t \de t' \int_0^{t-t'} \de \tau \Xjv^{irr}(t-t'-\tau) \gr
\Bigl( \Xfe(\tau) \eg(t')
+  \Xfv(\tau) \vv_0(t') \Bigr)
+ \\ &  +  \int_{-\infty}^t \de t' \Xjv^{irr}(t-t') \gr \Bigl( \Yfe \eg(t')
+ \Yfv \vv_0(t')
\Bigr).
\label{ams}
\end{split}
\end{equation}
\ii The third term can be simplified by first changing the variables of
integration to
$(t', t'' = t'+\tau)$, then changing the order of integration,
and finally using the fact that ({\it cf.} Eq.~\ref{wielkie3})
\begin{equation}
\int_{-\infty}^{t''} \de t' \Bigl( \Xfe(t''-t') \eg(t') +  \Xfv(t''-t')
\vv_0(t') \Bigr) = <\f>^{ret}_{t''}.
\end{equation}
By this means Eq.~\eqref{ams} can be rewritten as
\begin{equation}
\begin{split}
& <\jj>^{ret}_t  =  \int_{-\infty}^t \de t' (\Xje^{irr}(t-t') \eg(t') + \Xjv
^{irr}(t-t') \vv_0(t'))
+ \\ & \Yjv^{irr} \gr <\f>^{ret}_{t}  +
\int_{-\infty}^t \de t''  \Xjv^{irr}(t-t'') \gr <\f>_{t''}.
\label{ams2}
\end{split}
\end{equation}

The equations for the instantaneous and retarded current are then added
to yield the total current. The structure of the equations can be  most
clearly seen after the Fourier transform in time:
\be
<\jj(\omega)> =  <\jj(\omega)>^{inst}+<\jj(\omega)>^{ret} =
(\Yje^{irr}  + \Xje^{irr}(\omega)) \eg(\omega) + (\Yjv^{irr}+ \Xjv^{irr}(
\omega)) \vv(\omega)
\label{ms}
\ee
where we used the fact that the total suspension velocity may be written as
\be
<\vv>=\vv_0 + \gr <\f>
\ee
The Fourier transforms in time introduced above are defined as
  \begin{equation}
\vv(\omega) = \frac{1}{2\pi} \int_{-\infty}^{\infty} <\vv>_t e^{i \omega t}
\de t,
 \end{equation}
and analogously for $\jj(\omega)$, whereas the kernels $\X =<A e^{Lt} B>$  are
transformed as
  \begin{equation}
 \X(\omega) = \int_{0}^{\infty} <A e^{Lt} B> e^{i \omega t} \de t.
 \end{equation}

In an analogous way one may derive the equation for the average force
density (cf.~\ref{wielkie3}), getting

 \begin{equation}
\begin{split}
&  <\f(\omega)>  = (\Yfe^{irr}+\Xfe^{irr}) \eg(\omega) + (\Yfv^{irr}+\Xfe^{
irr}) <\vv(\omega)>
\label{j2a}
\end{split}
\end{equation}

The above result can be inserted into the Stokes  equation  to yield, after
the Fourier transform in space,

 \begin{equation}
\begin{split}
\eta k^2 <\vv(k,\omega)> = (\Yfe^{irr}+\Xfe^{irr}) \eg(\omega) + (\Yfv^{irr}
+\Xfv^{irr}) <\vv(k,\omega)>
\label{j2}
\end{split}
\end{equation}

\subsection{Small k expansions of response kernels}

In the long wave limit the tensor $\Yjv^{irr}$ takes
a particularly simple form. Namely, the scattering expansion \eqref{sce}
gives
 \be
\Yjv(\kb=0)^{irr} = \int  <\sum_{i=1}^N \delta(\R_i) [\sum_{l=0}^{\infty}
\pe^t  \mur_o \pe \ZZ_o (-
\G \z)^l]_{i}(\rr') >^{irr} \de \rr'
\ee
However, the integral over $\rr'$ may be replaced by the action of projection
operator $\pe^t$. Then, using Eq. \eqref{touse}, which implies that $\z \pe^t=0$,
we get
\be
\Yjv(\kb=0)^{irr} = <\sum_{i=1}^N \delta(\R_i)  \mur_o(i) \pe(i) \Z_o(i)
\pe^t(i)>^{irr} = <\sum_{i=1}^N \delta(\R_i)>
\boldsymbol{1} = n \boldsymbol{1}
\label{tamto}
\ee
where the expression \eqref{jedno} for one particle mobility matrix has been used.
The next nonvanishing term in the expansion of $\Yjv(\kb)$ in $\kb$ is the
second order one
\begin{equation}
\Yjv(\kb)^{irr} = n \boldsymbol{1} + k^2 \yjv+... ,
\end{equation}
\ii with the tensor $\yjv$ of the form
\begin{equation}
\yjv = y_{jv}^l \hkb \hkb + y_{jv}^t (\boldsymbol{1}- \hkb \hkb),
\end{equation}
\ii where $y_{jv}^l$ and $y_{jv}^t$ are scalars representing longitudinal and transverse
part of $\yjv$, respectively. Next, since $\Yfe^{irr}$ is adjoint to $\Yjv^{irr}$, we get
\begin{equation}
\Yfe(\kb=0)^{irr} = \Yjv(\kb=0)^{irr} = n \boldsymbol{1}.
\end{equation}
On the other hand, again using the property \eqref{touse} we get a simple
result
\begin{equation}
\Yfv(\kb=0)^{irr}=0
\end{equation}
\ii The small $\kb$ expansions of operators $\Yfe$ and $\Yfv$ read
 \begin{align}
 & \Yfe(\kb)^{irr} = n \boldsymbol{1} + k^2 \yfe +... , \no \\
 & \Yfv(\kb)^{irr} = k^2 \yfv +...
\end{align}
 Analogous expansions
are carried out for the time-dependent kernels
\begin{align}
 & \Xjv^{irr}(\kb,\omega) = k^2 \xjv(\omega) + \dots ,\label{xxje}  \\
 & \Xfv^{irr}(\kb,\omega) = k^2 \xjv(\omega)+ \dots , \\
 &  \Xfe^{irr}(\kb,\omega) = k^2 \xfe(\omega)+ \dots,
 \end{align}
 Using these expansions in Eqs. \eqref{ms} and \eqref{j2} one arrives at the
following relations for the diffusion current and force density for small
 but finite $\kb$:
\begin{subequations}
 \begin{align}
   \jj(\kb,\omega)-n \vv(\kb,\omega) =  \bigl( y_{jE} & +  x_{jE}(\omega)
\bigr)  \eg(\kb,
\omega) +  k^2 (
 \boldsymbol{1} - \hkb \hkb) \bigl( y_{jv}^{t} + x_{jv}^{t}(\omega) \bigr)
\VV(\kb,\omega), \no \\
 & \label{wk4} \\
  k^2 (\eta + y_{fv}^{t} + x_{fv}^{t}(\omega)) \vv(\kb,\omega) & = (
\boldsymbol{1} - \hkb \hkb) \Bigl(
 \f_0(\kb,\omega) + n \eg(\kb,\omega) \no \\
 & + k^2 \bigl( y_{fE}^{t} +  x_{fE}^{t}(
\omega) ) \eg(\kb,\omega)\Bigr)
 & \label{wk5}
  \end{align}
\end{subequations}
Here, $\yje$ and $\xje$ are defined by
\be
\Yje(\kb=0)^{irr} = \yje \boldsymbol{1}.
\label{yyje}
\ee
and
\be
\Xje^{irr}(\kb=0,\omega) = x_{jE}(\omega) {\bf 1}
\ee
Moreover, $y_{ab}^{t}$ and $x_{ab}^t$ denote the transverse part of the
operators $\boldsymbol{y}_{ab}$ and $\boldsymbol{x}_{ab}$ respectively and
the incompressibility
condition $(\boldsymbol{1} - \hkb \hkb) \vv(\kb) = \vv(\kb)$ was used.

The dynamics described by Eqs. \eqref{wk4} and \eqref{wk5} is relatively complex.
First of all, there are direct effects. First, an external force $\eg$ applied to the
particles induces the diffusion current
\be
\jj^d = \jj - n \vv
\ee
which is the particle current measured relative to the average suspension velocity frame.
The intensity of that effect is measured by the sedimentation coefficient, which now
becomes
frequency-dependent and reads
\be
K(\omega)=\frac{1}{n} (\yje+\xje(\omega))
\ee
Moreover, as seen in \eqref{wk5}, the suspension velocity field is induced by the overall
external force acting on
the particles and the fluid
\be
\F_{tot}=\f_0+n \eg
\ee
The effective viscosity of the suspension is modified by
the presence of the particles and reads
 \begin{equation}
\eta^{eff}(\omega) = \eta +y_{fv}^t+x_{fv}^t (\omega).
\end{equation}
Finally, there are also cross effects linking the suspension velocity $\vv(\omega)$ with
the external force acting on the particles $\eg(\omega)$ and the diffusion current
$\jj(\omega)$ with $\vv(\omega)$. The intensity of those couplings is measured by the
coefficients $y_{fE}^{t} +  x_{fE}^{t}(\omega)$ and $y_{jv}^{t} +  x_{jv}^{t}(\omega)$
respectively. However since $fE$ kernels are adjoint to $jv$ ones (cf. Eqs. \eqref{ker2d}
and \eqref{ker3}) the above coefficients are in fact equal. This is a manifestation of the
Onsager symmetry as suggested by Nozi{\`{e}}res \cite{Nozieres:1987}.

\section{Summary}

The response of a composite system with field-induced forces was studied using a newly
developed diagrammatic method. The method may be used in both instantaneous and retarded
response analysis. It was shown that in both cases it is possible to describe the system's
response by a set of transport coefficients which depend solely on local properties
of the medium. The expressions for the transport coefficients obtained with use of the
diagrammatic technique were shown to be well-behaved and free of divergences even in the
presence of long-ranged forces. Thus they represent a proper starting point for
calculation of the transport coefficients and for construction of approximate
methods.

As mentioned in the Introduction, a subsequent article~\cite{SadlejCichocki} will discuss
the application of the above methodology to the problem of the settling velocity and its
fluctuations in a non-Brownian suspension. This task is more complex than analogous
analysis for the Brownian suspension, presented in Sec.~\ref{sedim}, since the
distribution functions in that case correspond to the nonequilibrium (though stationary)
state. Using the diagrammatic technique one can derive correlation functions in this
state. Again, the crucial element of the derivation is the reduction procedure. Due
to its complexity, this procedure is nearly impossible to carry out  were it not for the
rigorous methodology provided by the diagrammatic method.

\appendix
\section{Simplification of time-dependent diagrams}

In this appendix we sketch the idea of the proof that the sum of all time-dependent
improper diagrams vanishes. The detailed proofs may be found in \cite{Szymczak:2001}.

First, let us introduce a few additional definitions concerning structure of the diagrams
from the expansion of $<A \evol{t} B>$. First, let us note that when one removes all
correlation functions and {\bf e}-bonds from a given diagram, it decomposes into a
number of subdiagrams - {\bf scattering blocks}, representing $A$, $B$ or subsequent
$\delta \boldsymbol{\cal L}$'s operators. The vertex in a given block which is most to the
left (right) will be called first (or last) vertex of the block respectively. Finally,
right(left) terminal block is a block with the property that the particle line
passing through its last(first) vertex {\it v} does not  pass through any other vertex in
a
diagram more to the right(left) than {\it v}.

For example, the diagram (D 6) consists of four scattering blocks. The first one (from the
left) is a left terminal $A$ block, the next one is a left terminal
$\delta \boldsymbol{\cal L}$ block. Then there is another $\delta \boldsymbol{\cal L}$
block and finally a right terminal $B$ block.

Note that every {\bf improper} diagram  must contain one of the following: either a right
(or left) terminal $\delta \boldsymbol{\cal L}$ block or a right terminal $A$ block
or a left terminal $B$ block. Next, we consider these cases in order.

The case when a diagram contains a right terminal $\delta \boldsymbol{\cal L}$ block is
relatively straightforward. It suffices to note that every $\delta \boldsymbol{\cal L}$
block ends with the \ \kq \ operator.
In the case of the right terminal $\dl$ block this
divergence operator has nothing to act on to its right and thus the value of such a
diagram
vanishes.

The case of left terminal $\dl$ block is a bit more complicated.
Let us denote such a block by $\boldsymbol{L_b}$. There are two possibilities:
\begin{description}
\item[a)] $\boldsymbol{L_b}$ begins with \ \kqq \ operator i.e.
$ \beta^{-1} \rabla_i \mur_o(i) \pe(i) \zo(i)$
\item[b)] $\boldsymbol{L_b}$ begins with $\F_{ji} \mur_o(i) \pe(i) \zo(i)$
\end{description}
Here $i$ denotes the particle with which $\boldsymbol{L_b}$ begins whereas $j$ is some
particle from the diagram different from $i$.
 For the diagrams in (a) , using integration by parts one can transform \ \kqq \
operator at the beginning of $\boldsymbol{L_b}$ for $-1 \cdot \ \lqq \ $ operator. But, as
$\boldsymbol{L_b}$ is the left terminal block, after such operation, the differentiation
in
${-1 \cdot} \ \lqq \ $ acts only on the correlation function on the far left of the
diagram. However, for the equilibrium distribution functions
\be
- \beta^{-1} \nabla_i n^{eq}_s(1,\dots,s) = - \sum_{j=1}^s {\bf F}_{ij} \
n^{eq}_s(1,\dots,
s)
- \int  {\bf F}_{i{(s+1)}}
n^{eq}_s(1,\dots,s+1) d(s+1)
\ee
Thus each diagram in (a) may be written as a sum of a number of diagrams in (b) with the
same scattering structure, taken with an opposite sign. In this way one can show that the
total sum of all diagrams in (a) and (b) vanishes.

It remains to consider two more cases: the
diagrams with a right terminal $A$-block and those with left terminal $B$-block. However,
in the first case, it suffices to transform \ $\llq$ operator at the end of $A$-block to
${-1 \cdot \ \kq}$ using integration by parts, and then note that again the divergence
operator has nothing to act on to its right.
Eventually, in case of diagrams with left terminal $B$-block,  the proof is analogous to
the one concerning left terminal $\dl$ blocks presented above

\section*{References}

\bibliographystyle{unsrt}
\bibliography{u1}


 \end{document}

%% file: defrys.tex
\newcommand{\kq}{\begin{picture}(5,5)
\thicklines
\put(-3,3){{ \circle{6}} }
\put(-3,6){{\small $\rightarrow$}}
\end{picture}}
\newcommand{\llq}{\begin{picture}(5,5)
\thicklines
\put(-3,3){{ \circle{6}} }
\put(-4,6){{\small $\leftarrow$}}
\end{picture}}
\newcommand{\kqq}{\begin{picture}(5,5)
\thicklines
\put(-3,3){{ \circle{6}} }
\put(-3,7){{\tiny $\Rightarrow$}}
\end{picture}}
\newcommand{\lqq}{\begin{picture}(5,5)
\thicklines
\put(-3,3){{ \circle{6}} }
\put(-4,7){{\tiny $\Leftarrow$}}
\end{picture}}

\newcommand{\sztylet}{\begin{picture}(10,10)
\multiput(0,-5)(0,6){3}{$\boldsymbol{\dagger}$}
\end{picture}}

\newcommand{\kd}{\begin{picture}(10,10)
\put(-2,0){{\thicklines \circle{10} }}
\end{picture}}
\newcommand{\kdp}{\begin{picture}(10,10)
\put(-2,0){{\thicklines \circle{10} }}
\end{picture}}
\newcommand{\klr}{\begin{picture}(10,10)
\put(-2,0){{\thicklines \circle{10} }}
\thicklines
\put(-8,9){{$\Rightarrow$} }
\end{picture}}
\newcommand{\kpr}{\begin{picture}(10,10)
\put(-2,0){{\thicklines \circle{10} }}
\thicklines
\put(-7,9){{$\rightarrow$} }
\end{picture}}
\newcommand{\kppl}{\begin{picture}(10,10)
\put(-2,0){{\thicklines \circle{10} }}
\thicklines
\put(-5,9){{$\leftarrow$} }
\end{picture}}

\newcommand{\akd}{\begin{picture}(10,10)
\put(5,5){{\thicklines \circle{10} }}
\end{picture}}

\newcommand{\kore}{\begin{picture}(10,10)
\put(-1,0){{ \bf{$\bigcap $}} }
\put(-1,-40){{ \bf{$\bigcup $}} }
\thicklines
\put(4.0,0){\line(0,-1){34}}
\put(11.8,0){\line(0,-1){34}}
\end{picture}}

%% file: defsym.tex
\newcommand{\here}{{}}

\newcommand{\ii}{\noindent}

\newcommand{\Le}{\boldsymbol{{\cal L}}}

\newcommand{\seso}{S(0)}

\newcommand{\h}{h}

\newcommand{\dl}{\delta \boldsymbol{\cal L}}
\newcommand{\de}{\mbox{d}}
\newcommand{\be}{\begin{equation}}
\newcommand{\ee}{\end{equation}}
\newcommand{\ba}{\begin{eqnarray}}
\newcommand{\ea}{\end{eqnarray}}

\newcommand{\vv}{{\boldsymbol{v}}}
\newcommand{\ub}{{\boldsymbol{u}}}

\newcommand{\mur}{\boldsymbol{\mu}}
\newcommand{\zet}{\boldsymbol{\zeta}}
\newcommand{\VV}{{\boldsymbol{V}}}

\newcommand{\jj}{{\boldsymbol{j}}}

\newcommand{\Y}{{\boldsymbol{Y}}}
\newcommand{\Yje}{{\boldsymbol{Y}}_{jE}}
\newcommand{\yfv}{{\boldsymbol{y}}_{fv}}

\newcommand{\yjv}{{\boldsymbol{y}}_{jv}}
\newcommand{\yje}{y_{jE}}
\newcommand{\Yjv}{{\boldsymbol{Y}}_{jv}}

\newcommand{\Yfe}{{\boldsymbol{Y}}_{fE}}
\newcommand{\yfe}{{\boldsymbol{y}}_{fE}}
\newcommand{\Yfv}{{\boldsymbol{Y}}_{fv}}

\newcommand{\Xje}{{\boldsymbol{X}}_{jE}}
\newcommand{\xje}{x_{jE}}

\newcommand{\xjv}{{\boldsymbol{x}}_{jv}}
\newcommand{\Xjv}{{\boldsymbol{X}}_{jv}}

\newcommand{\Xfe}{{\boldsymbol{X}}_{fE}}
\newcommand{\xfe}{{\boldsymbol{x}}_{fE}}
\newcommand{\Xfv}{{\boldsymbol{X}}_{fv}}

\newcommand{\evol}[1]{e^{\boldsymbol{\cal L}#1}}

\newcommand{\labla}{\stackrel{\leftarrow}{\nabla}}
\newcommand{\rabla}{\stackrel{\rightarrow}{\nabla}}

\newcommand{\D}{{\boldsymbol{D}}}
\newcommand{\E}{\boldsymbol{\cal E}}
\newcommand{\G}{\boldsymbol{\cal G}}
\newcommand{\gr}{{\boldsymbol{G}}}
\newcommand{\M}{{\boldsymbol{M}}}

\newcommand{\Z}{{\boldsymbol{Z}}}

\newcommand{\T}{{\boldsymbol{T}}}

\newcommand{\U}{{\boldsymbol{U}}}
\newcommand{\X}{{\boldsymbol{X}}}

\newcommand{\R}{{\boldsymbol{R}}}
\newcommand{\rr}{{\boldsymbol{r}}}
\newcommand{\rrh}{\Hat{\boldsymbol{r}}}
\newcommand{\w}{{\boldsymbol{\Omega}}}

\newcommand{\zzr}{\boldsymbol{\Hat{\cal Z}}}

\newcommand{\zr}{{\boldsymbol{\Hat{Z}_o}}}
\newcommand{\z}{{\boldsymbol{\Hat{\cal Z}_o}}}
\newcommand{\zo}{{\boldsymbol{Z_o}}}

\newcommand{\ZZ}{\boldsymbol{\cal Z}}
\newcommand{\pe}{\boldsymbol{\cal P}}

\newcommand{\eg}{{\boldsymbol{E}}}

\newcommand{\kb}{{\boldsymbol{k}}}
\newcommand{\hkb}{\Hat{{\boldsymbol{k}}}}

\newcommand{\F}{{\boldsymbol{F}}}

\newcommand{\pot}{\frac{\partial}{\partial t}}
\newcommand{\pox}{\frac{\partial}{\partial {\boldsymbol{X}}}}

\newcommand{\mo}{\mu_o}

\newcommand{\C}{{\boldsymbol{\tilde{\cal C}}}}
\newcommand{\Ce}{\boldsymbol{\cal C}}

\newcommand{\f}{{\boldsymbol{f}}}
\newcommand{\no}{\nonumber}